# Influence Maximization (IM) in Complex Networks with Limited Visibility Using Statistical Methods

Saeid Ghafouri[1], Seyed Hossein Khasteh[2], Seyed Omid Azarkasb[3]✉

[1] MEng of Artificial Intelligence, K.N. Toosi University of Technology, Tehran, Iran
Saeidghafouri@email.kntu.ac.ir

[2] Assistant Professor of Artificial Intelligence, K.N. Toosi University of Technology, Tehran, Iran
Khasteh@kntu.ac.ir

[3] Ph.D. Student of Artificial Intelligence, K.N. Toosi University of Technology, Tehran, Iran
Seyedomid.azarkasb@email.kntu.ac.ir

**Abstract:**

A social network (SN) is a social structure consisting of a group representing the interaction between them. SNs have recently been widely used and, subsequently, have become suitable and popular platforms for product promotion and information diffusion. People in an SN directly influence each other's interests and behavior. One of the most important problems in SNs is to find people who can have the maximum influence on other nodes in the network in a cascade manner if they are chosen as the seed nodes of a network diffusion scenario. Influential diffusers are people who, if they are chosen as the seed set in a publishing issue in the network, that network will have the most people who have learned about that diffused entity. This is a well-known problem in literature known as influence maximization (IM) problem. Although it has been proven that this is an NP-complete problem and does not have a solution in polynomial time, it has been argued that it has the properties of sub modular functions and, therefore, can be solved using a greedy algorithm. Most of the methods proposed to improve this complexity are based on the assumption that the entire graph is visible. However, this assumption does not hold for many real-world graphs. This study is conducted to extend current maximization methods with link prediction techniques to pseudo-visibility graphs. To this end, a graph generation method called the exponential random graph model (ERGM) is used for link prediction. The proposed method is tested using the data from the Snap dataset of Stanford University. According to the experimental tests, the proposed method is efficient on real-world graphs.

## 1- Introduction

A social network (SN) is made up of a group of social activities with nodes between them. The overall structure of SNs was shown in Figure 1. SNs are web-based application sites that allow users to communicate with friends and family, meet new people or friends, join groups, chat, share photos, organize events, or create a network with people who have similarities with them in their daily lives. An important problem in any real-world network is to identify influential nodes in that network and define them as selecting a set of people in the network so that it has the maximum influence over the people of the network and causes a wide spread of the diffusion process (informing and creating culture). The nodes that need more attention and investment to perform a specific task can be found by identifying the influential people in the network from the perspective of various parameters. For example, what should be the priority of displaying pages in web browsers? How are the nature and extent of conspiracy involved in criminology identified? What are the most important points in a biological network? Where did an infectious disease spread from? Which politicians are the most influential in a network of politicians? Different people are not equally important. All these questions show the serious necessity of finding the influential nodes of a network. The goal is not only to detect the influential nodes in many cases but also to detect the nodes with the maximum influence in diffusing an entity in the network. Assume a company with a limited budget intends to promote a product in an SN through a marketing campaign. The ideal method for this purpose is for this company to reach each member of this network directly and make them aware of this product. However, such a method will not be practical due to the limited budget. The obvious solution is for the company to ask only a limited number of influential network members to promote its product. This has led to the emergence of a research branch called influence maximization (IM) [1].

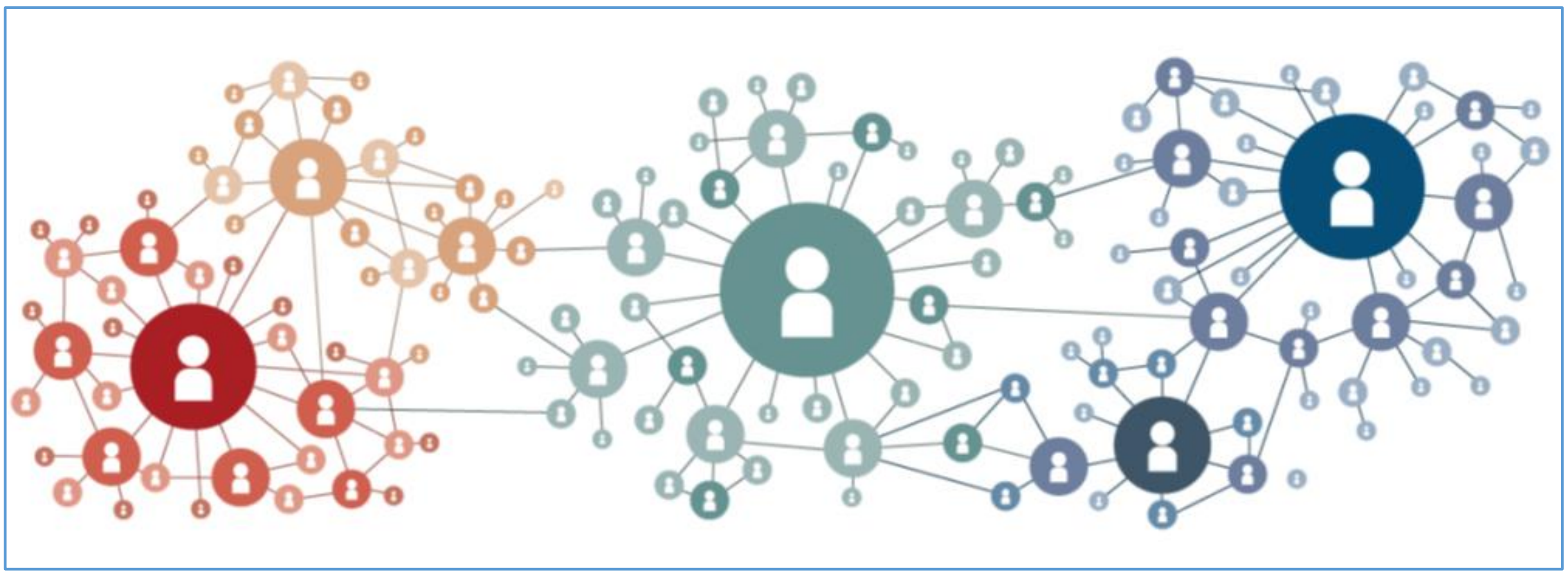

Figure 1. The overall structure of SNs

Missing data in SNs is another major challenge in these networks. In other words, the data extracted from SNs have connections that are not present in our dataset for several reasons, including sampling error. In another example, SN graph data can be accessed only if you are a premium member of the SN provider. The inaccuracy of the data has led to the development of a research topic as "link prediction" [2-4]. Meanwhile, various centrality measures are proposed, which can be divided into radial and medial. The random step in the radial state starts from or ends at one node. However, in the medial state, the random step passes through the node. The radial measure is divided into long-based and volume-based depending on the type of random step selected. The long-based measures set the volume of the target nodes, trying to find the step length to achieve the target volume. On the other hand, the volume-based measure does the opposite. Centrality measures include degree centrality, closeness centrality, betweenness centrality, and structural holes. Degree centrality is a centrality measure based on the radial-volume measure and the easiest centrality measure because it simply indicates the degree of a node. This measure is suitable for measuring the total number of connections of a node but does not necessarily always indicate the importance of a node. Here, the matrix can be used along with the network to calculate the degree of centrality because all its values have a length of one and start from the node. Furthermore, the centrality of path length k which counts the number of paths with maximum length k can be used [4]. Closeness centrality is a centrality measure based on the radial-length measure. Closeness centrality indicates how close a node is to all other nodes in the network. It is calculated as the average of the shortest path length from the node to every other node in the network. If the centrality value is small, the node is central. Betweenness centrality is a type of medial centrality. The high value of this measure indicates the basic position of the edge. On the other hand, the high value indicates the conditional position of the node, that is, it allows the diffusion of a lot of information. The betweenness centrality is calculated based on the existence of the node in the closest path of a pair of nodes, just like the betweenness centrality of the edge of the node. The measure indicates the importance of the node in the network in the field of information diffusion. In the meantime, the structural nodes are defined as connecting nodes and create several local bridges in the network. The removal of these nodes developed spaces in the network, which are referred to as structural holes. The node creating the structural hole can connect the information from the non-interactive nodes. So, this node has a significant contribution to connecting different areas of the network [5]. This study is generally conducted to find influential nodes or to solve the IM problem in networks with limited visibility. In this regard, the IM methods are combined with the link prediction techniques. Influential nodes have maximum influence if they have a higher centrality score. The rest of this manuscript is prearranged as follows: Section 2 depicts a brief review of recent research related to our proposed technique and is presented in two categories. The diagram of the proposed method and dependent algorithms are presented in section 3. Section 4 gives the technical considerations of implementation and analysis of results. Finally, the conclusion and future studies are summed up in section 5.

## 2- Literature Review

The members of an SN share different topics and diffuse different information among themselves. In these networks, one person's opinions, thoughts, and beliefs can change those of other members. So, SN analysts have paid a lot of attention to expanding influence and identifying influential people. One of the most important problems in influence optimization is to find a small subset of social people so that, by activating them, the information of the largest number of people from the network is influenced under a diffusion model. In [1] on IM, solving the problem of finding influential diffusers was defined for the first time in the form of an optimization problem, and it was found that it is an NP-complete problem. However, it was argued that it is a submodular problem under certain conditions. Accordingly, it can be solved using a greedy algorithm with an approximation close to the optimal solution. According to Equation 1, the function f is submodular if and only if the value added in the function f due to adding a member c to the set A is greater than or equal to the same value if it is added to the set B for both sets A and B such that A⊆B.

$$f(A \cup \{c\}) - f(A) \geq f(B \cup \{c\}) \qquad \text{(Eq.1)}$$

The theorem proposed in [6] suggests that submodular functions can be calculated by a greedy solution with an approximation of $1 - 1/e$ such that, if the optimal solution of the function is $f(S^*)$, the solution $f(S)$ of the greedy method will be $f(S) \geq \left(1 - {}^{1}/_{e}\right). f(S^*)$. However, this approximation is also true for maximizing the dynamic influence [7].

A large number of studies have been published every year, in each of which the initial algorithm has been improved in some ways [8-12]. You can refer to the review papers presented in this field such as [13] and [14] for more information. The initial algorithm has been improved in one of the following cases in subsequent studies. For example:

- Review of different diffusion models [15-19];
- Examining different greedy algorithm models [9], [11], [20];

- Using the capabilities of different heuristic methods to reduce the execution time and required memory [16], [21-23];
- Examining different networks from different perspectives [24-26].

In the following, other related methods presented in this field are reviewed in more detail and their advantages and noteworthy issues are mentioned in the subsections of identifying the influential nodes and influence maximization (IM) for more familiarity with the subject. It is worth noting that the noteworthy issues have been extracted from the section on future studies and a detailed examination of their text by the authors of the present study.

### 2.1 Identifying the Influencing Nodes

As mentioned earlier, the identification of influential people includes the selection of a group of people in the network with the maximum influence over the other people in the network and causes a wide spread of the diffusion process. Besides, removing inactive nodes can provide a sparser graph. This sparse graph makes a great contribution to the speed of social media influence and analysis. In the following, some of the recently proposed methods to find and identify influential nodes in SNs are reviewed in order of publication year. [27] identified the influential nodes in online SNs by using an accurate nearest neighbor (NN) imputation algorithm. The proposed method takes into account the position of neighboring and non-neighboring nodes and the delay in the diffused information. The advantage of the method is its compatibility with the SN application environment. In [28], a new ranking method with several features was used to find the position of a node and its neighboring nodes. The proposed method has advantages such as low complexity and suitability for large scale. In this study, the entropy method was used to determine the position of a node and the features of neighboring nodes. This method improved the accuracy of finding the influencing node. The authors used the position measure to identify the positional difference between the removed nodes. The number of repetitions in the method was an average of 50 executions. Influential nodes in Twitter were identified based on network topology and user behavior in [29]. After the development and modeling of local networks, the next step of the personalized ranking algorithm is to model users with specific user features to analyze the local influence. One of the indicators for evaluating the performance of the personalized ranking algorithm is the number of tweets of nodes influencing the identification of their information diffusion potential. The advantage of the presented method is the correct use of user behavior to calculate the influence of each user. One of the noteworthy issues of the method is the high overhead due to the dynamic nature of the network. This method is used together with a distributed method to accelerate calculations. According to the results of a test on a large dataset collected from Twitter, it is more effective to use user-specific features and network features to identify influential nodes in a specific topic. In this study, various tests were conducted to compare the evaluation indices that are usually used to evaluate social influencing nodes. The experimental results suggest that a simple evaluation index or measure, the diffusion rank, is effective in verifying the influential nodes. The features and behavior of users, which are the same nodes, and an international business network application were used to analyze SNs in [30]. Facilitating and completing users' understanding of situational changes in countries caused by the participation of edge weights in the calculation of indicators is one of the advantages of this method. In this study, the international trade network was analyzed by contributing its structures and the features defined for the node, and the scatter plots between the calculated indicators were presented in different ways. [31] provided a scalable method to find influential nodes in SN community detection. One of the noteworthy issues about this method is that it does not allow overlap, and its advantage is not needing to have prior knowledge of the number of influential sets and nodes. In the proposed method, the influential nodes are found first and the communities are then detected. A similarity function is considered to calculate the proximity of the active node with the highest centrality score. Two assumptions are taken into account. The first one is that a node can only belong to a set. The second one is the sharing of members of a set consisting of a large number of neighbors with one influential node. A supervised scalable statistical method was used in [32] to identify influential nodes in online SNs and remote users were identified and classified into seven different classes. The advantage of the proposed method is the accuracy of classification in seven different classes. High overhead due to a large number of classes is one of the noteworthy issues of the method. The authors in [33] identified influential nodes in SNs based on neighborhood diversity and proposed new methods, DSR and EDSR, to rank influential nodes. The advantages of the proposed method are the higher accuracy and validity and the separation of nodes according to their influence. The computational overhead due to the use of two algorithms and the lack of implementation on large datasets are noteworthy issues of the proposed method. In this method, an index was presented to determine the centrality of the node using Shannon entropy and Jensen-Shannon divergence. In this index, the influence of a node was determined based on the distribution of its neighbors in the graph, the space of their influence in the network, and the intensity of the influence. In [34], an efficient method was proposed to identify influential nodes in dynamic and scalable networks using a local detection and updating strategy. The presented method has advantages such as higher speed for small and random networks, working on a multi-core processor to increase speed for scale-free networks, and predicting changes in the influence of specified communities in dynamic networks. The increase in computational overhead is one of the noteworthy issues of the proposed method. In this study, the goal of IM was implicitly followed along with the identification of influential nodes. [35] identified influential nodes in SNs with the approach of voting from the neighbors of each node. According to the results of tests and simulations using the SIR model in many real datasets, the proposed method was better than some common methods. In this study, a voting-based method was presented to find the influencing nodes. In this method, a set of influential nodes is selected based on a voting scheme in which each node has the same voting abilities and receives votes from its neighbors. However, the authors argued that each node should have different voting abilities depending on its topological position in the network. One of the advantages of the proposed method is the improvement of accuracy, and one of the noteworthy issues is increasing overhead due to the voting of each node separately. [36] was conducted to identify influential nodes in SNs based on correlation coefficients. It was found that the number of common neighbors with a node and its neighbors determine the influence of that node. In the presented method, a clustering approach is used in which the common hierarchy of nodes and their neighborhood set are considered. Some tests have been carried out on artificial and real networks to show the effectiveness of the proposed ranking approach. Considering the neighborhood parameter and improving the accuracy of finding influential nodes are the advantages of the proposed method. [37] was conducted to identify the influential nodes based on the user's activity and behavior in online SNs. Identification of influential nodes by analyzing online SNs can effectively reach the target audience. So, it has attracted the attention of researchers. Although user behavior has a significant contribution to increasing the maximum influence, it is not considered in most studies. In the proposed method, a model is provided to identify influential nodes by evaluating the influence factors of members based on their activity in online SNs. The advantage of the method is to consider the user's behavior and activity, and one of the noteworthy issues is not using artificial intelligence to predict the user's behavior and activity. In the following, the IM methods will be investigated. However, some of these methods are introduced here because they are focused on identifying influential nodes. Other methods, for example, in [38-41], will be discussed in more detail in the next subsection. An optimal method was presented in [38] to select influential nodes in SNs. To do this, a new and simple IM model was presented, which could be used to investigate the diffusion of short messages through mobile phone-based SNs. The provided model was multi-objective integer programming (MOIP). The advantage of this method is its simplicity in simulation and execution. The influence rate was calculated in [39] to maximize the influence in dynamic networks. In this method, the upper bound interchange was presented by the algorithm to solve the problem based on tracking the influential node using the greedy interchange method. The advantage of the presented method is its low execution time, and one of the noteworthy issues is not using the upper limit interchange to follow the influential nodes and not generalizing the linear threshold model (LTM) in dynamic networks. The method aims to follow the influencing nodes dynamically to maximize the influence. A new method based on the gradient method was proposed in [40] to maximize the influence of diffusion. The advantage of the method is to maximize the influence of diffusion and optimize the execution time. This algorithm ensures the best minimum diffusion through the initially selected nodes. However, one of the noteworthy issues is that the greedy algorithm only obtains the effect of local minimum propagation in each iteration. So, it cannot access more nodes. The results suggested that different business strategies can be planned by identifying the influential nodes and the very high rate of influence of diffusion in social media. This method used the dynamic threshold. In [41], a two-stage selection algorithm was presented to increase the maximum influence in SNs. The problem was to maximize the influence to find a small subset of nodes in a social activity. In the presented method, a certain number of influential nodes were selected as candidate nodes. In the first stage, a certain number of nodes were assigned as candidate nodes. In the second stage, the maximum performance was performed to estimate the effect of each candidate node. Finally, the nodes were selected from the candidate nodes according to their maximum influence. Implementation in six social networks and improving accuracy and efficiency are the advantages of this method. One of the noteworthy issues is increasing the execution time. A method to identify influential nodes in social networks was introduced in [42] using the community structure and the difference in influence distribution. For this, in the first stage, a network embedding-based community detection approach was developed, by which the SN was divided into several high-quality communities. This stage is referred to as the candidate stage. The candidate stage consists of selecting candidate nodes from inside and the border of each community using a heuristic algorithm. In the second stage, the greedy method based on submodular features was used to select influential nodes with maximum marginal influence from the candidate set. According to the experimental results, the proposed method can ensure the expansion of the influence and the reduction of execution time in SNs on a large scale compared to the existing methods. One of the noteworthy issues in this method is to consider the influence of the content of the node on the spread of influence. Moreover, node classification and sentiment analysis methods can be used to identify influential nodes. Since user behavior and social link play an important role in information diffusion, this method can consider these factors for nodes. Influential nodes in complex networks were identified in [43] through a distance-based effective centrality mechanism. The proposed algorithm considered things like K-shell strength, node degree, effective distance, several levels of neighbor influence, or neighborhood potential. Thus, this algorithm can be applied to any network, i.e., directed or undirected. The performance of the proposed algorithm was evaluated on nine real-world networks, where a modified susceptible-infected-recovered (SIR) model was used to investigate the spread dynamics of each node. In [44], it was discussed that some existing influential node detection algorithms do not consider the influence of edges, and, consequently, the algorithm's influence deviates from the expected limit. Some consider the global structure of the network, which leads to high computational complexity. According to the simulation results, the proposed algorithm performed better than other existing techniques such as betweenness centrality, eigenvector centrality, closeness centrality, hyperlink-induced topic search, page rank, profit leader, H-index, K-shell, and gravity over a valuable margin. One of the noteworthy issues in this method is adding more parameters to adjust the intensity of the influence. A method to identify influential nodes using the local structural features of the network according to the information entropy theory was proposed in [44] to solve the above problems. The influential nodes were evaluated based on the entropy and weight distribution of the edges connected to them to calculate the difference in the weight of the edges and the influence of the weight of the edges on the neighboring nodes. In this method, the accuracy of ranking the influence of nodes should still be improved because this method only focuses on the influence of the first and second-order edges of the nodes. In [45], the influential nodes were identified by combining the centrality measures using symbolic regression (SR), which can identify suitable mathematical expressions that fit the network features to combine these measures. The new mathematical expressions perform better in the ranking than the Pearson correlation, Jaccard similarity score, and Kendall's Tau-b correlation. One of the noteworthy issues in this method is the examination of suitable measures that can be combined. The methods of identifying the influencing nodes are summarized in Table 1.

Table 1. The methods of identifying the influencing nodes

| Reference Number | Year of Publication | Method | Advantages | Noteworthy Issues |
| --- | --- | --- | --- | --- |
| [38] | 2016 | An optimal method to select influential nodes in SNs | The simplicity of simulation and execution | Not applicable for large volumes of data |
| [27] | 2017 | Identifying influential nodes in online SNs by accurately evaluating neighbors | Suitability with SN application environment | No overlap |
| [28] | 2017 | Finding the position of a node and neighboring nodes with a new ranking | Low complexity and suitability for a large scale | Not using more robust optimizations to increase accuracy |
| [39] | 2017 | Calculating the influence rate and tracking the influential nodes in dynamic networks | Low execution time | Not using upper bound interchange and generalization of linear threshold model (LTM) |
| [29] | 2018 | Identification of influential nodes in Twitter based on network topology and user behavior | Proper use of users' behavior to calculate the influence of each user | High overhead due to the dynamic nature of the network |
| [30] | 2018 | Analysis of SNs using the characteristics and behavior of users (nodes) | Users' understanding of situational changes | Increase in overhead due to network dynamics and changing the position of nodes |
| [31] | 2018 | A scalable method to find influential nodes in SNs | No need for prior knowledge of the number of sets and active nodes | No overlap |
| [32] | 2018 | A scalable method to find influential nodes in SNs | The accuracy of classification in seven different classes | High overhead due to a large number of classes |
| [33] | 2019 | Identifying influential nodes in SNs based on neighborhood diversity | Improving the accuracy and validity and separating the nodes according to their influence | Computational overhead and failure to examine large datasets |
| [34] | 2019 | Identifying influential nodes in dynamic and scalable networks using local detection and updating strategy | Speed up for small scale-free networks and prediction of changes in dynamic networks | Increased computational overhead |
| [40] | 2019 | A gradient-based method to maximize the effect of diffusion | Optimizing execution time and ensuring the best diffusion through initial node selection | Lack of access to more nodes |
| [35] | 2020 | Identifying influential nodes in SNs by voting from the neighbors of each node | Improving accuracy | Increased overhead due to the voting from each node separately |
| [36] | 2020 | Identifying influential nodes in SNs based on correlation coefficients | Considering the neighborhood parameter and improving the accuracy of finding active nodes | Not using the centrality measure of each node |
| [37] | 2020 | Identifying influential nodes based on user activity and behavior in online SNs | Considering user behavior and activity | Not using artificial intelligence to predict user behavior |
| [41] | 2020 | An algorithm for selecting influential nodes to increase the maximum impact in SNs | Implementation in six social networks and improving accuracy and efficiency | Increasing execution time |
| [42] | 2021 | Identifying influential nodes in SNs through community structure | Achieving an effective balance between the spread of | Not examining the influence of node content, user |
| | | and the difference in influence distribution | influence and execution time | behavior, and social connection on information diffusion |
| [43] | 2021 | Identifying influential nodes in complex networks through a distance-based effective centrality mechanism | Implementation in nine social networks and improved performance compared to other existing techniques | Adding more parameters to adjust the intensity of the influence |
| [44] | 2022 | Identifying influential nodes using local structural features of the network | Examining eight real-world networks with different network structural features | Not paying attention to the influence of third and higher-order edges of nodes |
| [45] | 2022 | Identifying influential nodes by combining centrality measures using symbolic regression | Better performance than the latest indices such as Pearson correlation, Jaccard similarity score, and Kendall's Tau-b correlation | Examining suitable measures that can be combined |

### 2.2 Influence Maximization

To the best of the authors' knowledge, readers are well acquainted with the concept of influence maximization (IM). In the following, studies on IM will be reviewed.

According to [46], limited information diffusion is an NP-hard problem whose solution requires high time complexity. Traditional greedy algorithm or heuristic algorithm is computationally expensive in large SNs. The problem studied in the above study is the limit of final influence, which is derived from the problem of limiting the spread of false information in social networks and proposes an independent cascade model that models the diffusion by two cascades with simultaneous evolution in the network. The advantage of the presented method is not to spread false information in SNs. One of the noteworthy issues is not considering the model based on finding the active node for information diffusion. The authors investigated the influence blocking maximization problem for the first time under the competitive linear threshold model, which was an extension of the classical model, in [47], arguing that the objective function of the influence blocking maximization is uniform under the competitive linear threshold model. In this study, the influence blocking maximization problem was defined and its submodularity was proved under the competitive linear threshold model. Besides, the optimal computation feature of the model was used for directed non-modular graphs, and an optimal heuristic method was proposed for the influence blocking maximization problem under the competitive linear threshold model to increase the computational gain. Examining the influence blocking maximization for the first time is the advantage of the proposed method. One of the noteworthy issues is not examining the limitations and challenges of influence blocking maximization. The least-cost rumor blocking (LCRB) problem in SNs was investigated in [48]. In this study, an algorithm was proposed to limit the bad influence. The algorithm initiated a support cascade for diffusion against the rumor cascade. Accordingly, limiting the bad influence is the advantage of the proposed method. One of the noteworthy issues is the lack of implementation on different datasets. The rumor blocking problem in online SNs was investigated and a model to minimize the influence of rumors on user experience was proposed in [49]. In the proposed blocking strategy, the influence of blocking time on user experience in online social networks was considered. The advantage of the proposed method is to consider the influence of the rumor blocking time. One of the noteworthy issues is dependent on previous information. In [50], the authors investigated the influence blocking maximization problem and proposed two heuristic strategies to solve the problem optimally. The main idea of this method is to find a group of active people who start the diffusion of good information, thus maximizing the influence of blocking the diffusion of bad information in online SNs. Finding active nodes to diffuse good information is the advantage of the proposed method. One of the noteworthy issues of the proposed method is not using a more robust optimization to find active nodes. In [51], an attempt was made to solve the IM problem in SNs with community structure, and the label diffusion protocol was introduced to identify influential nodes in SNs. The labels were obtained from some core nodes, and the centrality of the core node was evaluated based on the label diffusion process. The influential nodes were determined to maximize the influence after ranking the core nodes according to the centrality. The advantage of the method is to take into account the label diffusion process for each node because one node may belong to several real social network communities. One of the noteworthy issues of the method is the lack of improvement in finding active nodes through more robust optimization algorithms. The influence was maximized by activating the link in SNs in [52]. Many studies have recently been conducted on the diffusion of innovations in SNs. Previous studies mostly focused on IM by identifying a set of early adopters or on influence minimization by blocking links under a specific diffusion model. This method considered the IM problem by activating the link with the independent cascade model. An approximate solution was presented for this problem by calculating the cost degree coefficient for selecting the active link. The advantage of the proposed method is to get better results on the real network. Some noteworthy issues of the method are the non-generalization of the model and the linking of the parameters. The authors in [53] investigated the minimum cost influence diffusion in SNs. They provided a new optimization problem and generalize a variety of previous problems inspired by viral marketing in SNs. In this method, precise and heuristic algorithms were presented based on an ILP formulation with exponential variables that allowed the inclusion of optional activation functions and strengthened valid inequalities. According to the computational results, the proposed approaches performed significantly better than the existing algorithms and their extensions in specific cases of the general problem. More robust optimization compared to other methods is the advantage of the proposed method. One of the noteworthy issues of the method is the lack of improvement of the heuristic separation routines and the lack of identification of more inequalities. In [54], the improvement of multi-objective IM in SNs was investigated. In the presented method, the multi-objective evolutionary approach was improved for IM in SNs. This method finds the set of k central nodes that maximizes the nodes obtained in the network because the minimization of the value of k is also presented as an optimization objective. The main disadvantage of the previous evolutionary approach was the time required to achieve good solutions. However, the proposed method showed how the initialization of the first generation leads to better convergence and a Pareto optimal profile. This approach was applied to three real-world SNs. The problem of predicting participants in the diffusion of information in social networks and its applications were discussed in [55]. The most important available information is the approximation of the adoption probability of users, which models the behaviors and propensity to diffuse in the six categories of presented features. Simultaneous use of early adopters and other users with the highest estimated adoption probabilities can yield satisfactory results. The advantage of this method is to provide more effective targeted marketing, i.e., not tracking active nodes to improve information diffusion in SNs. In [56], the authors investigated influence diffusion for social graph-based recommendations and introduced an influence diffusion algorithm with a threshold to determine this cascade influence. They defined three conditions to determine the threshold of a node for influence and used three approaches to initially rank the nodes. They then evaluated these variables with experimental analyzes on real-world datasets. The results showed that node-dependent threshold conditions performed better than global threshold conditions. In this study, the proposed algorithm was used to generate social graph-based neighborhoods. These graphs were considered as input to the algorithm. One of the noteworthy issues of this method is not considering other centrality measures, influence diffusion strategies, and different thresholding to further improve the process. The advantage of the method is the use of influence diffusion as an important feature for information diffusion. [57] introduced a new influence diffusion model to maximize influence based on membership in SNs and included membership marketing in the study of the IM problem for the first time. In this study, a multi-stage influence diffusion model was developed based on membership marketing features. The developed model divided the influence diffusion process into two stages, influence and referral. So, the proposed algorithm measured the capability of each node in the influence and referral stages, ranked the nodes based on the weighted sum of the capabilities of the two stages, and continuously selected the node with the highest weighted sum as the central node. In [58], the authors investigated the effect of awareness on IM in SNs and applied the presented method to solve other problems, for example, to maximize the influence of conscious distance. Extensive tests on real and synthetic data showed the effectiveness, efficiency, and scalability of the method. According to the results obtained from real big data, this method could have improved by 58% on average compared to competitors. The advantage of the proposed method is its scalability, and one of its noteworthy issues is the high computational overhead. In [59], useful information was extracted from several levels of neighbors for a target group to estimate its influence strength. In this method, a centrality measure was presented, which estimated the influence power of a node by summing the weights of its multi-level neighbors, which were mainly determined based on their distance to the target node. The advantage of the proposed method is its excellent performance and stability on nine networks with different sizes and categories. The proposed centrality measure was a superior alternative centrality that was more appropriate and accurate than degree centrality. Not using multi-objective optimizations to solve multiple functions is one of the noteworthy issues in the provided method. IM was investigated in [60] based on the proximity of communities in SNs. In the presented method, the algorithm for increasing the maximum influence based on the proximity of the community was proposed to select the influencing nodes and the point-to-point influence of the community was reflected. The advantage of the method is to improve the results on synthetic and real data, and one of its noteworthy issues is not using optimization methods to select influential nodes. In [61], the authors proposed a new algorithm for IM in complex networks with community structure without the need to determine the number of seed nodes. The proposed algorithm identified influential nodes with three methods in each stage (degree centrality, random hole, and structural hole) in each community and measured the spread of influence again in each stage. This process continued until the end of the algorithm. Finally, the most influential nodes with maximum diffusion in each community were identified. The community-based detection approach enabled the algorithm to find more influential nodes than those suggested by page rank and other centrality measures. According to [62], IM algorithms mainly focus on one-to-one influence diffusion among users with friendships. However, in addition to one-to-one friendships, there are usually one-to-many group links in real social settings that have rarely been fully considered by conventional methods. In this study, a truncated meta-seed generator was presented to select a small number of users based on the two components of friendships and group links. Furthermore, a structural seed developer was proposed to extend the meta-seed set to encode distinct diffusion structures of friendships and group links. In general, a good balance was established between effectiveness and efficiency to improve performance. One of the noteworthy issues is the non-presentation of used pseudo-codes. An improved IM method in SNs was presented in [63], which is one of the most recent studies in this field. In this regard, soft computing such as a dynamic generalized genetic algorithm was used in SNs under independent cascade models to obtain a dynamic influential set. In this study, several graphs were modeled in changing edges and nodes in different time frames, which led to effective changes in the number of members of the seed set. Reduction of computing costs and maintenance of optimization process in dynamic SNs are the advantages of this method. Non-implementation in the real environment is one of the noteworthy issues in this method. The IM problem was defined as a pseudo-regression in [64] using several deep learning and machine learning techniques. The idea behind this study was embedded in a graph and using a neural network. An embedded struc2vec node was used to create an embedded for each node in the network based on the idea of exploiting the structural identity of nodes in a network in the initial phase of the algorithm. These node embeddings then served as feature vectors for the network nodes. The messaging system of graph neural networks (GNNs) was then used, and the generated node embeddings were passed to the regression. The influence of each node in the training network was calculated using the information diffusion model and formed labels for regression while training the model. The trained model was used to predict the possible influence on the test networks using regression. Finally, a set of size k was selected by selecting the top k nodes based on their predicted influence. LSTM cells were used as the neighborhood pooling function for the artificial network part of the graph, trying to optimize the error between the calculated and predicted influence for the regression part. The proposed method was compared and evaluated with the SIR model and the independent cascade model. One of the noteworthy issues in this method is the lack of comparison with more models. The IM methods are summarized in Table 2.

Table 2. IM methods

| Reference Number | Year of Publication | Method | Advantages | Noteworthy Issues |
|---|---|---|---|---|
| [46] | 2011 | Investigating the diffusion of false information in SNs | No diffusion of false information in SNs | Not examining the model based on finding effective nodes for information diffusion |
| [47] | 2012 | The IBM problem under the competitive linear threshold model in SNs | Investigating the influence blocking maximization for the first time | Not examining the limitations and challenges of IBM |
| [48] | 2013 | The least-cost rumor blocking (LCRB) problem in SNs | Limiting the bad influence | Lack of implementation on different datasets |
| [49] | 2016 | Solving the IM problem in SNs with community structure | Considering more than one diffusion process label for a node | Not improving influential nodes through more robust optimization algorithms |
| [50] | 2017 | The rumor blocking problem in online SNs | Considering the influence of the time of the rumor blocking | Dependence on previous information |
| [51] | 2017 | The influence blocking maximization (IBS) problem in SNs | Finding influential nodes to diffuse good information | Not using a more robust optimization to find influential nodes |
| [52] | 2018 | IM by activating links in SNs | Better results on the real network | Non-generalization of the model and linking of the parameters |
| [53] | 2018 | The least-cost influence diffusion in SNs | More robust optimization than other methods | Not improving heuristic separation and not identifying further disparities |
| [54] | 2018 | Improving the evolutionary multi-objective IM in SNs | Better performance than the new heuristic methods | No increase in speed and no checking of larger networks |
| [55] | 2018 | The problem of predicting participants in the information diffusion in SNs | Providing more effective targeted marketing | Not tracking influential nodes to improve information diffusion |
| [56] | 2018 | Influence diffusion for social graph-based recommendations | The use of influence diffusion as an important feature | Not examining other centrality measures and various thresholds |
| [57] | 2019 | A new influence diffusion model for IM based on membership in SNs | Analysis of the activity and proximity of the nodes in different periods | Not investigating more robust algorithms to quantify the effectiveness |
| [58] | 2020 | The effect of awareness of IM in SNs | Scalability | Computational overhead |
| [59] | 2020 | An approach to IM in SNs | Excellent performance and stability on nine networks of different sizes | Not using multi-objective optimization to solve multiple functions |
| [60] | 2020 | IM based on the proximity of communities in SNs | Improved results on synthetic and real data | Not using optimization methods to select influential nodes |
| [61] | 2021 | IM in complex networks based on community structure | IM without the need to determine the number of seed nodes and special attention to degree | Lack of examination of other centrality measures and their |
| | | | centrality, random hole, and structural hole | influence on the diffusion process |
| [62] | 2021 | IM in multi-relational SNs | Improving effectiveness and efficiency by considering a truncated meta-seed generator and including friendships and group links | No presentation of pseudo-codes used |
| [63] | 2022 | Improving IM in SNs based on soft computing such as genetic algorithm | High scalability, reduction of computing costs, and maintenance of the optimization process in dynamic SNs | Lack of implementation in real environments |
| [64] | 2022 | Transforming the IM problem in complex networks into a pseudo-regression problem | Better performance than classical methods | No comparison with more than two models |

Although many studies have been conducted on IM, it can be claimed that most of them assume access to the entire graph structure. However, this assumption is unrealistic in many cases. This study tries to expand the capabilities of the existing IM methods to graphs that we do not have a complete view of. In other words, the existing graph has many unseen edges in many cases. In these cases, specific algorithms must be designed for these conditions or existing algorithms must be manipulated so that they can be extended to new conditions. In this study, the second approach was used. A probabilistic graph generation model called the exponential random graph model (ERGM) was used to achieve this goal. ERGM, which was first introduced in [65], aims to develop a probability distribution of the graph based on the frequent substructures in the graph. This probability distribution can be used in the next steps to create graphs with the same properties as the original graph. The frequent substructures include the number of any type of substructure of the graph, such as the number of triangles, edges, and more complex substructures. This model of graphs was generated using graph generation capabilities. In the next step, the graphs with added edges were used as inputs to the IM algorithms.

According to [66] and [67], ERGMs were introduced as a new research field for the first time in [65]. Although these graphs were called Markov graphs at that time, they were the same concepts and properties that were referred to as exponential random graph models in future studies. In each ERGM, a probability is assigned to each producible graph. In other words, a probability distribution is defined for each possible graph on a specific set of nodes. Two important concepts in these models are the number of frequent substructures and the parameters assigned to each of them. Each of these substructures is a specific framework of graph nodes and edges that are repeated throughout the graph, such as triangles, edges, or paths of a certain length. ERGM aims to provide a probability distribution of graphs with similar properties by using this number of sets of these subgraphs and assigning a coefficient to each of them. [65] was the first study to suggest that these substructures can serve as sufficient statistics for a log-linear model. Sufficient statistics for a model indicate a case where adding more complexity to the model cannot add more power to it. So, ERGMs are representations of a set of graphs based on their frequent sub-recurrences. One of the stages of developing the ERGM model is matching the relevant parameters and finding the appropriate coefficients. Various machine learning methods have been used for this purpose, most of which are based on the maximum likelihood estimation (MLE). Refer to [68] for more information.

## 3- The Proposed Method

In this section, the proposed method is introduced step by step. In this method, a link prediction step was performed before the execution of the IM algorithm to add the ability to operate graphs with limited visibility to the current IM methods. Figure 2 shows the steps of this method.

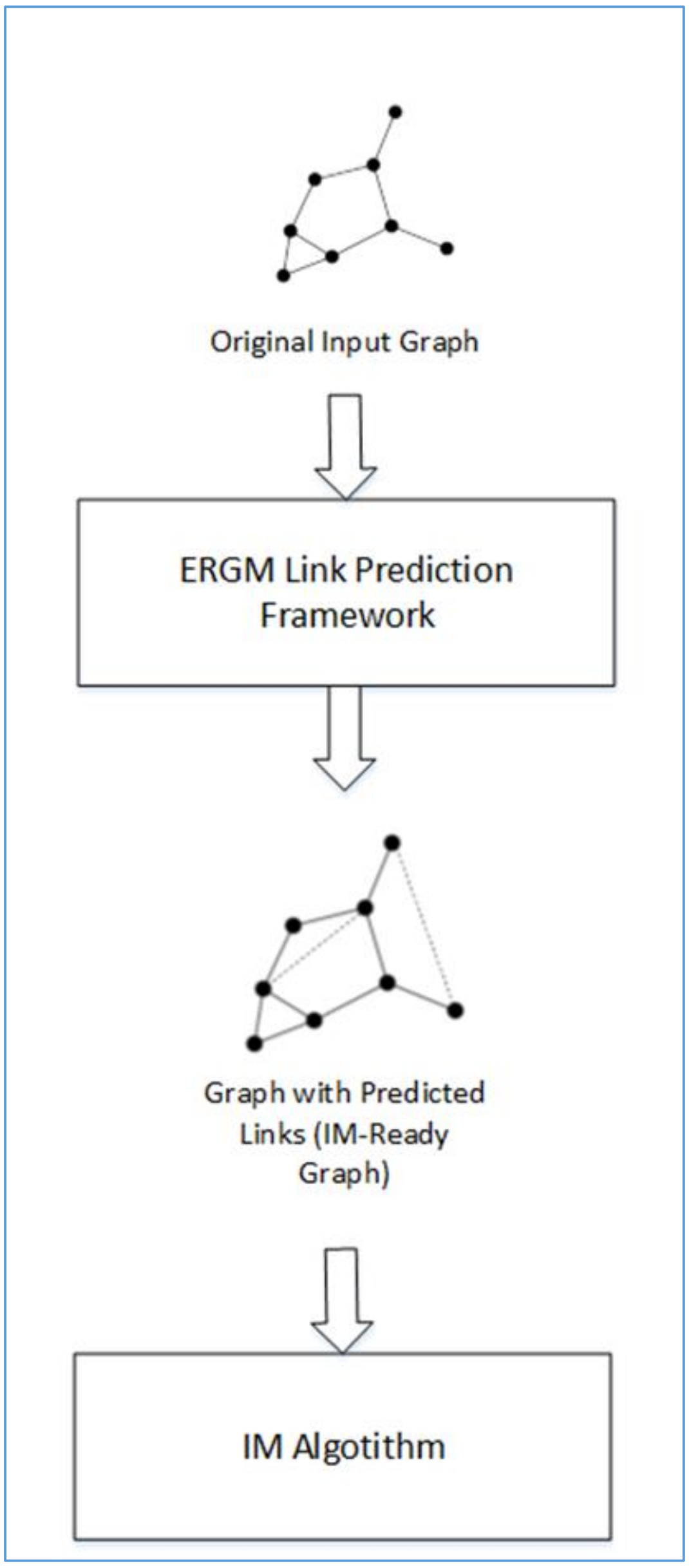

Figure 2. The general steps of the proposed method

The symbols used in this section are given in Table 3.

Table 3. The symbols used to introduce the proposed method

| Sign | Features |
|---|---|
| $S$ | The seed set of the IM problem |
| $I$ | The diffusion model of the IM problem |
| $\Gamma(S)$ | The diffusion function specifies the number of activated nodes |
| $\sigma(S)$ | The mean diffusion function |
| $IM(G.k.I)$ | The IM function |
| $G$ | The original graph input to the link prediction method |
| $A$ | The graph with added edges or the output of the link prediction method |
| $M$ | The set of medial graphs generated in the link prediction method |
| $m_i$ | An arbitrary medial graph |
| $n$ | The number of generated medial graphs |
| $L(G)$ | The link prediction method used |
| $i$ | The index of the generated graphs |
| $e$ | The probability of a graph edge |
| $\theta$ | The threshold to separate predicted edges |
| $tr(\theta.d_i)$ | The function separating useful edges from link prediction |

The primary output of the proposed method for predicting the link of a graph is a probability on each edge that indicates the presence of that edge in the original graph. According to Equation 2, the link prediction method L returns graph A with added edges in exchange for the word graph G. Each of the added edges $e_i$ has a probability in the range of 0 to 1.

$$L(G) = A \quad \text{(Eq.2)}$$

The noteworthy issue is that a large number of these added edges are very unlikely to exist. So, they are very unlikely to exist in the original graph. According to Equation 3, a simple method to extract useful edges is to determine a threshold θ for the edges and remove the edges that are less than this threshold. Another method is to sort the edges based on the probability of their existence and select some edges with the highest probability.

$$tr(\theta . d_i) = \begin{cases} if\ d_i \geq \theta & Do\ not\ trim \\ if\ d_i < \theta & tirm \end{cases} \quad \text{(Eq.3)}$$

In this study, the framework that was first introduced in [3] was used for link prediction. The graph with unseen edges was first fitted on an ERGM. Many graphs were then generated from this ERGM. Here, these graphs are referred to as medial graphs. In this step, there is a set of synthetic graphs $M = \{m_1 . m_2 \ldots\ldots m_n\}$. In an adjacency matrix, the number of repetitions of each edge in the set of all generated graphs is kept. Each entry of this adjacency matrix is then divided by the total number of sets of generated graphs, i.e., n. So, the probability of each arbitrary edge $e_{i.j}$ is calculated according to Equation 4.

$$e_{i,j} = \left(\sum_{k=1}^{n} m_k\right) / n = \sum_{k=1}^{n} \left(\sum_{e \in m_k} e_{m_{k\,i,j}}\right) / n \quad \text{(Eq.4)}$$

Here, the generated graphs are averaged. After this step, there is an adjacency matrix in which each entry is the probability of a specific edge in the original graph. The framework of this method can be seen in Figure 3. This framework was previously mentioned by the authors of the present study in [69], which we have expanded in this article and present as future works of our previous article.

According to the above, Algorithm 1 is introduced as follows:

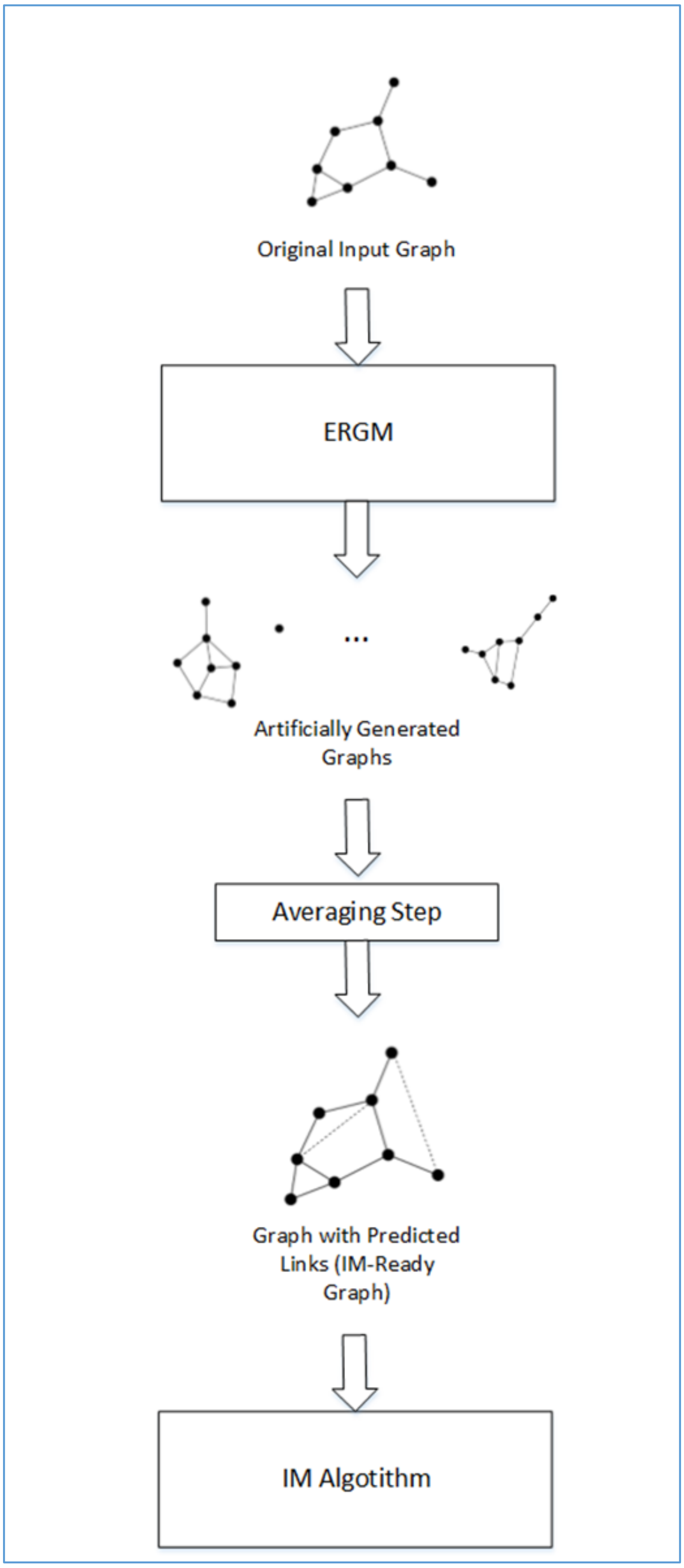

Figure 3. The general framework of the link prediction method using ERGM [69]

According to the above, Algorithm 1 is introduced as follows:

**Algorithm 1:** Link Predicted GreedyIM

**Input:** Graph $G = (V, E, W)$, $k$, diffusion model $I$, trimming value $\theta$, original diffusion values for each edge $p_i$

$L(G) = A$ # depends on the used link prediction framework
**for each edge** $e_i$ **in** $A$
  **if** $e_i < \theta$: # any other trimming criteria can also be used
    **delete** $e_i$ **from** $A$
  **end if**
**end for**
$G \leftarrow A$
$S \leftarrow \emptyset$
$i \leftarrow 0$
**while** $(i < k)$ **do**
  $i \leftarrow i + 1$
  $v^* \leftarrow \text{argmax}_{\forall v \in V}\{\sigma(S \cup \{v\}) - \sigma(S)\}$ under $I$
  $S \leftarrow S \cup \{v^*\}$
**end while**
**Return** $S$

Algorithm 1: The pseudo-code of the proposed algorithm

The link prediction method can be used for other types of link prediction methods, except for the proposed method based on ERGM, before the execution of the IM algorithms. Therefore, the more general framework can be expressed as Algorithm 2.

**Algorithm 2:** Link Predicted General IM

**Input:** Graph $G = (V, E, W)$, $k$, diffusion model $I$, trimming value $\theta$, original diffusion values for each edge $p_i$

$L(G) = A$ # depends on the used link prediction framework
**for each edge** $e_i$ **in** $A$
    **if** $e_i < \theta$: # any other trimming criteria can also be used
        **delete** $e_i$ **from** $A$
    **end if**
**end for**
$IM(G, k, I) = S$ # depends on the used IM framework
**Return** $S$

Algorithm 2. The pseudo-code of the framework more general than the proposed algorithm

### 3.1 Implementation Details

In this section, the method used during the performance evaluation tests, all the details of the dataset used, and the specifications of the computing environment used are described in detail.

#### 3.1.1 Programming Languages and Libraries Used

In this study, several languages and technologies were used to implement the proposed method, each of which is briefly introduced. The data were preprocessed using Python version 3. The NetworkX library [70] was used for the implementation of the graphic parts. It is currently the most famous Python language library for graph processing. Besides, the famous Numpy and Pandas libraries, which are the most important libraries for data pre-processing and post-processing, were widely used. The parts concerning ERGM were implemented using the R language and the well-known library for statistical processing of graphical data called Statnet [71]. The best implementation of ERGM among all existing libraries is used in the Statnet library. Since the innovation of this study was only the operationalization of the previous IM methods in environments with limited visibility, the previous IM methods were implemented for the tests. The IMrank [7] and Static greedy [8] methods were used during the tests. Moreover, the codes provided by the authors of these studies, which were in C++ language, were used in the implementations. All implementations are available at https://github.com/sdghafouri/IMinPO.

### 3.1.2 The Data Used

In this study, the following two datasets were used to check the performance of the proposed method.

- Data from the General Relativity and Quantum Cosmology Collaboration Network
- Data from the High Energy-Physics Theory Collaboration Network

The above data were provided from the Snap dataset of Stanford University, which is freely available to the public at http://snap.stanford.edu. Both these datasets were first presented in [72].

These data were collected from General Relativity and Quantum Cosmology Collaboration Network published on the arXiv website. Each node in the graph represents one of the authors in the entire dataset. Each undirected edge represents a collaboration between two authors on at least one paper. This dataset covered the papers from January 1993 to April 2003 (124 months in total). Tables 4 and 5 show the specifications of the graph composed of these datasets.

Table 4. The data from the first dataset

| | |
|---|---|
| Nodes | 5242 |
| Edges | 14496 |
| Nodes in the largest WCC | 4158 (0.793) |
| Edges in the largest WCC | 13428 (0.926) |
| Nodes in the largest SCC | 4158 (0.793) |
| Edges in the largest SCC | 13428 (0.926) |
| Average clustering coefficient | 0.5296 |
| Number of triangles | 48260 |
| Fraction of closed triangles | 0.3619 |
| Diameter (longest shortest path) | 17 |
| 90-percentile effective diameter | 7.6 |

Table 4. The data from the Second dataset

| | |
|---|---|
| Nodes | 9877 |
| Edges | 25998 |
| Nodes in the largest WCC | 8638 (0.875) |
| Edges in the largest WCC | 24827 (0.955) |
| Nodes in the largest SCC | 8638 (0.875) |
| Edges in the largest SCC | 24827 (0.955) |
| Average clustering coefficient | 0.4714 |
| Number of triangles | 28339 |
| Fraction of closed triangles | 0.1168 |
| Diameter (longest shortest path) | 17 |
| 90-percentile effective diameter | 7.4 |

### 3.1.3 Computing Resources

In this study, several computing machines with different specifications were used for tests, the specifications of each of which are briefly given in Table 6.

Table 6. The specifications of computing machines

| **First Machin** | |
|---|---|
| RAM | 16 Gb |
| CPU Model | Core™ i7-4702MQ (6MB Cache, up to 3.20GHz) |
| GPU | No separate GPU |
| Operating System | Ubuntu 18.04 |
| **Second Machin** | |
| RAM | 16 Gb |
| CPU Model | Core™ i7-7700K (8MB Cache, up to 4.20GHz) |
| GPU | Nvidia Geforce™ GTX 1080 Ti |
| Operating System | Windows™ 10 Enterprise |
| **Third Machin** | |
| RAM | 64 Gb |
| CPU Model | Core™ i7-7700K (8MB Cache, up to 4.20GHz) |
| GPU | Nvidia Geforce™ GTX 1080 Ti |
| Operating System | Windows™ 10 Enterprise |

### 3.1.4 Performance Evaluation Method

Some edges of the graph were first removed randomly. The number of edges removed was set by a predefined variable indicating the percentage of edges to be removed. In this step, two other graphs called "added" and "random" graphs were formed from the graph with removed edges. The "added" graph was generated by adding the equal number of edges that were removed from the original graph in the previous step. These new edges were generated by the link prediction method described in the previous section. The same number of edges was added to the random graph by randomly selecting some edges from the set of all possible edges. Up to this point, there are three graphs: 1) the original graph, 2) the added graph (with edges added through link prediction, and 3) the random graph with randomly added edges. In the next step, the IM algorithm was executed on each of the graphs, and, subsequently, a diffusion step was performed on the graph. In the output of this step, three sets were obtained:

O: the infected nodes in the original graph

A: The infected nodes in the added graph

R: The infected nodes in the random graph

The sharing between nodes in each of the sets was calculated through Equations 5 and 6.

$$b = |O \cap R| \quad \text{(Eq.5)}$$

$$c = |O \cap A| \quad \text{(Eq.6)}$$

The added graph was expected to have more infected shared nodes compared to the original graph. So, the value of c was greater than b. The measure used to show the superiority of the proposed method was comparing the difference between c and b. It is worth mentioning that the graphs R and A differed only in a part of the edges. For example, if 20% of the edges were removed from the original graph, only 20% of the edges would be different and 80% of the graphs would be similar in the later steps (however, the nodes would be completely matched). The difference between c and b was divided by the difference between the graphs for a fair comparison. For example, if two graphs differed only in d% edges, the result of the difference between c and b had to be divided by this value because the only difference between these two graphs was in d% edges and it was quite expected that 100-d% of the graphs would have nodes infected in the original graph. The measure $M_1$, $\mathrm{M}_2$, and $M_3$ were introduced for evaluation. The measures are shown in Equations 7, 8, and 9.

$$M_1 = \left(\frac{c-b}{t}\right) * \left(\frac{1}{d\%}\right) \quad \text{(Eq.7)}$$

In Equation 7, t represents all the infected nodes in the original graph or the size of the set $O$ or $|O|$.

$$M_2 = \left(\frac{c-b}{b}\right) * \left(\frac{1}{d\%}\right) \qquad \text{(Eq.8)}$$

In the trends shown in Figures 5-10, the measure $M_3$ was used.

$$M_3 = (c-b) * \left(\frac{1}{d\%}\right) \qquad \text{(Eq.9)}$$

In the ERGM in this study, the following frequent subgraphs were used.

- Edges: The number of edges in the input graph
- Isolated nodes: the number of nodes with zero degree
- Geometrically weighted degree count: refer to [73].
- Geometrically weighted dayadwide count: refer to [73].
- Geometrically weighted edgewise count: refer to [73].

In the next step, 1000 graphs were developed using this model, which was matched to the desired graph. These graphs were then averaged. The edges in the original graph had a probability of one regardless of the output of the link prediction method.

In this study, two robust IM methods were used. The first one was IMrank [7], which is a method based on unifying the steps of calculating the influence function and selecting the seed set. The second one was Static Greedy [8], which is a method that preserves the subscale property of the diffusion function in addition to improving the execution time. In addition to the IM methods, two centrality-based methods called betweenness centrality [74] and page rank [75] were used for comparison. Betweenness centrality measures the number of times a node lies on the shortest path between other nodes and determines the influence of communication flows. A node with a higher betweenness degree has a position in the network that can link other pairs or groups in the network. Page rank is a technique that is based on probability distribution and shows how likely a user is to click on a link to reach the desired web page. In this technique, the rank of each page is calculated based on the rank of the pages that refer to it. Page rank is useful in determining which node (web page) has the most significant contribution among different nodes. A clear example of the concept of page rank is shown in Figure 4.

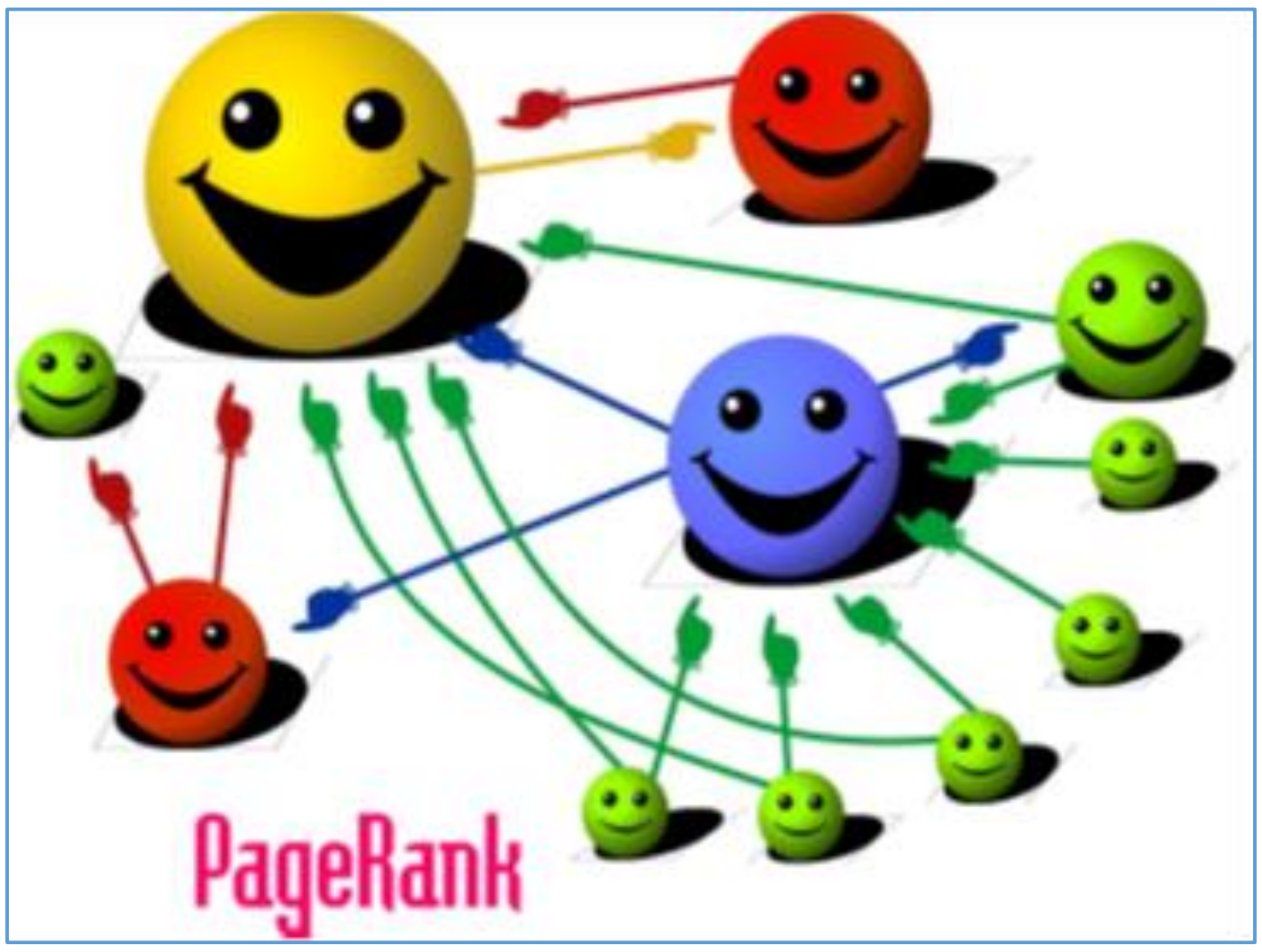

Figure 4. The concept of page rank

Except for the seed set obtained from the algorithm, the seed set obtained from random selection was also investigated. After selecting the seed set by each of these methods, a diffusion test was performed on them and each of the original, added, and random graphs. The seed set size in all tests was 100. Each diffusion test was performed 100 times on each of the graphs to avoid the influence of the randomness of the results. The obtained results were the average of 100 diffusion tests.

### 3.1.5 Test Measures and Variables

The values presented in Table 7 were used for the variables in the tests. Dynamic variables are those whose values are tested for several different values, and static variables have the same value throughout all tests.

Table 7. The parameters used in the tests

| **link prediction variables** | | | |
|---|---|---|---|
| **Parameter** | **Description** | **Value** | **Type** |
| $f$ | The similarity between the graphs | 0.9, 0.85, 0.8, 0.75, 0.7 | Dynamic |
| $numOfSimNets$ | The number of generated medial graphs | 1000 | Constant |
| **Diffusion variables** | | | |
| **Parameter** | **Description** | **Value** | **Type** |
| $SizeOfSeed$ | The size of the IM seed set | 100 | Constant |
| $diffP$ | The diffusion probability used in the tests | 0.25, 0.2, 0.15 | Dynamic |
| $R$ | The static greedy algorithm variable | 200 | Constant |
| $L$ | The IMrank algorithm variable | 1 | Constant |
| $iters$ | The IMrank algorithm variable | 10 | Constant |
| **Influence Maximization variables** | | | |
| **Parameter** | **Description** | **Value** | **Type** |
| $numOfSims$ | The number of diffusion tests | 100 | Constant |
| $numOfIters$ | The number of diffusion steps | 100 | Constant |

## 4- Implementation and Analysis of Results

The results of the tests are given in Tables 8-13. The method column indicates the maximization measure used, and the added and random columns indicate the sharing of the number of active nodes after diffusion with the active nodes of the original graph in each of the corresponding graphs. The total column represents the active nodes in the original graph.

### 4.1 The Results of the First Dataset

Table 8. The results of the first dataset with a diffusion probability of 0.25

| **Diffusion Probability = 0.25** | | | | **Similarity Percentage = 0.9** | |
|---|---|---|---|---|---|
| **Method** | **Added** | **Random** | **Total** | **M1** | **M2** |
| Betweenness Centrality | 1084.11 | 1071.14 | 1736.6 | 7.46 | 12.10 |
| PageRank Centrality | 1054.69 | 1053.5 | 1703.52 | 0.69 | 1.12 |
| IM Rank | 1165.25 | 1150.43 | 1808.78 | 8.19 | 12.88 |
| Static Greedy | 1197.09 | 1179.77 | 1861.78 | 9.30 | 14.68 |
| Random | 1184.28 | 1175.05 | 1831.27 | 5.04 | 7.85 |
| **Diffusion Probability = 0.25** | | | | **Similarity Percentage = 0.85** | |
| **Method** | **Added** | **Random** | **Total** | **M1** | **M2** |
| Betweenness Centrality | 1098.15 | 1079.56 | 1734.08 | 7.14 | 11.47 |
| PageRank Centrality | 1068.96 | 1045.12 | 1714.5 | 9.26 | 15.20 |
| IM Rank | 1155.26 | 1140.98 | 1811.35 | 5.25 | 8.34 |
| Static Greedy | 1188.1 | 1170.24 | 1851.37 | 6.43 | 10.17 |
| Random | 1146.73 | 1138.15 | 1789.5 | 3.19 | 5.02 |
| **Diffusion Probability = 0.25** | | | | **Similarity Percentage = 0.8** | |
| **Method** | **Added** | **Random** | **Total** | **M1** | **M2** |
| Betweenness Centrality | 1097.79 | 1085.7 | 1742.76 | 3.46 | 5.56 |
| PageRank Centrality | 1059.17 | 1053.75 | 1707.05 | 1.58 | 2.57 |
| IM Rank | 1148.43 | 1135.01 | 1803.21 | 3.72 | 5.91 |
| Static Greedy | 1198.43 | 1178.38 | 1860.2 | 5.38 | 8.50 |
| Random | 1153.72 | 1148.06 | 1806.08 | 1.56 | 2.46 |
| **Diffusion Probability = 0.25** | | | | **Similarity Percentage = 0.75** | |
| **Method** | **Added** | **Random** | **Total** | **M1** | **M2** |
| Betweenness Centrality | 1097.88 | 1083.72 | 1721.64 | 3.28 | 5.22 |
| PageRank Centrality | 1068.74 | 1054.48 | 1707.97 | 3.33 | 5.40 |
| IM Rank | 1161.39 | 1149.4 | 1808.25 | 2.65 | 4.17 |
| Static Greedy | 1199.16 | 1186.5 | 1868.61 | 2.71 | 4.26 |
| Random | 1168.71 | 1159.25 | 1811.64 | 2.08 | 3.26 |
| **Diffusion Probability = 0.25** | | | | **Similarity Percentage = 0.7** | |
| **Method** | **Added** | **Random** | **Total** | **M1** | **M2** |
| Betweenness Centrality | 1108.8 | 1091.83 | 1732.86 | 3.26 | 5.18 |
| PageRank Centrality | 1071.01 | 1051.89 | 1715.22 | 3.71 | 6.05 |
| IM Rank | 1161.83 | 1148.43 | 1810.43 | 2.46 | 3.88 |
| Static Greedy | 1194.12 | 1181.97 | 1847.06 | 2.19 | 3.42 |
| Random | 1171 | 1156.54 | 1797.44 | 2.68 | 4.16 |

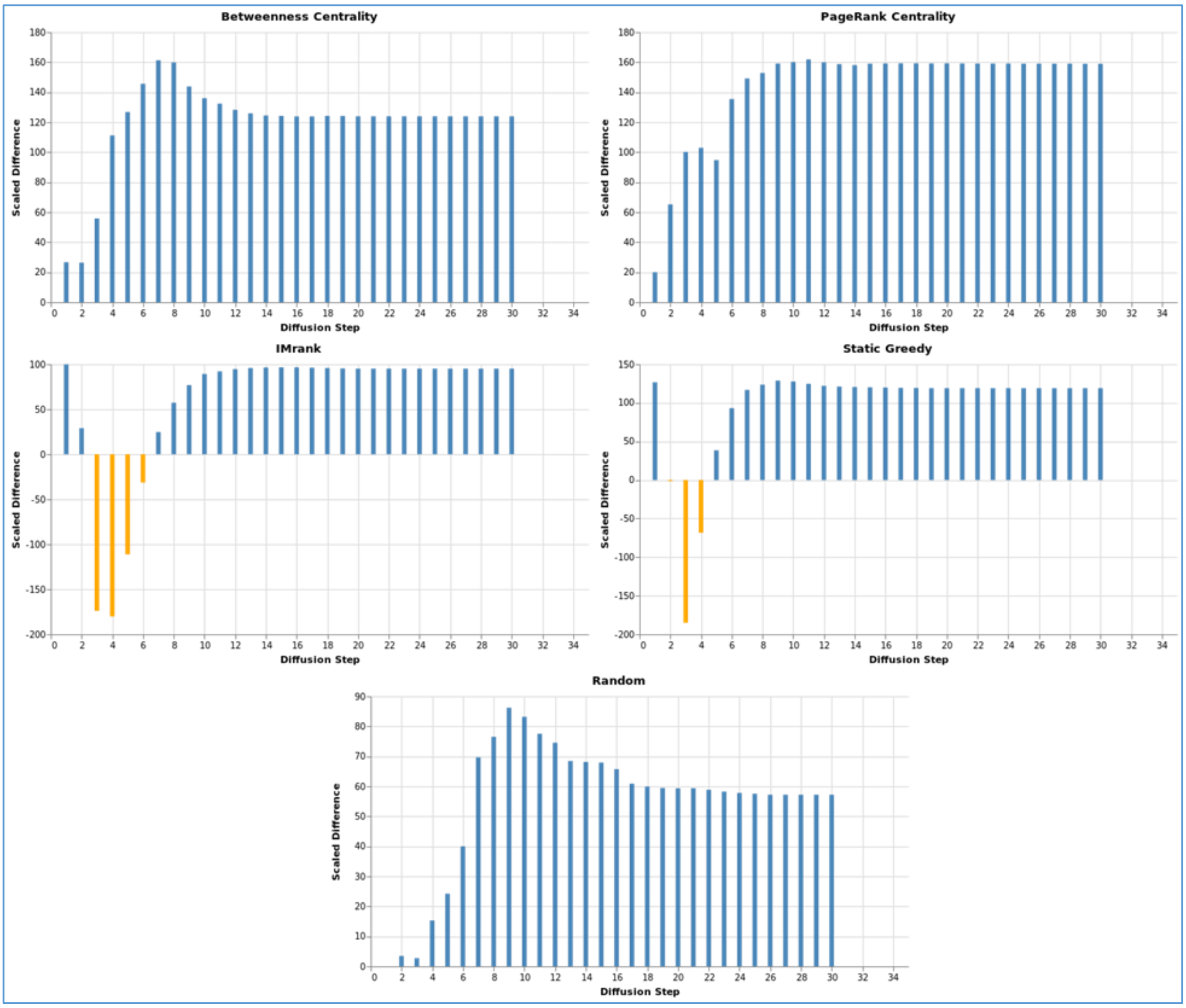

Figure 5. The trend at a similarity of 0.85 and a diffusion probability of 0.25

Table 9. The results of the first dataset with a diffusion probability of 0.2

| **Diffusion Probability = 0.2** | **Similarity Percentage = 0.9** |
|---|---|

| Method | Added | Random | Total | M1 | M2 |
|---|---|---|---|---|---|
| Betweenness Centrality | 740.93 | 721.41 | 1341.66 | 14.54 | 27.05 |
| PageRank Centrality | 703.74 | 679.01 | 1321.72 | 18.71 | 36.42 |
| IM Rank | 785.37 | 784.73 | 1418.75 | 0.45 | 0.81 |
| Static Greedy | 835.29 | 819.67 | 1493.35 | 10.45 | 19.05 |
| Random | 778.82 | 764.83 | 1381.2 | 10.12 | 18.29 |
| **Diffusion Probability = 0.2** | | | | **Similarity Percentage = 0.85** | |
| **Method** | **Added** | **Random** | **Total** | **M1** | **M2** |
| Betweenness Centrality | 749.94 | 721.25 | 1353.19 | 14.13 | 26.51 |
| PageRank Centrality | 702.25 | 673.04 | 1314.57 | 14.81 | 28.93 |
| IM Rank | 783.28 | 776.7 | 1423.47 | 3.08 | 5.64 |
| Static Greedy | 831.93 | 808.35 | 1493.96 | 10.52 | 19.44 |
| Random | 776.3 | 745.21 | 1367.68 | 15.15 | 27.81 |
| **Diffusion Probability = 0.2** | | | | **Similarity Percentage = 0.8** | |
| **Method** | **Added** | **Random** | **Total** | **M1** | **M2** |
| Betweenness Centrality | 722.58 | 706.54 | 1342.04 | 5.97 | 11.35 |
| PageRank Centrality | 686.28 | 660.56 | 1328.71 | 9.67 | 19.46 |
| IM Rank | 763.5 | 756.25 | 1420.38 | 2.55 | 4.79 |
| Static Greedy | 811.27 | 792.85 | 1498.66 | 6.14 | 11.61 |
| Random | 769.48 | 747.64 | 1386.59 | 7.87 | 14.60 |
| **Diffusion Probability = 0.2** | | | | **Similarity Percentage = 0.75** | |
| **Method** | **Added** | **Random** | **Total** | **M1** | **M2** |
| Betweenness Centrality | 742.98 | 694.17 | 1357.27 | 14.38 | 28.12 |
| PageRank Centrality | 683.51 | 643.04 | 1323.32 | 12.23 | 25.17 |
| IM Rank | 779.67 | 741.52 | 1418.67 | 10.75 | 20.57 |
| Static Greedy | 824.65 | 779.04 | 1510.41 | 12.07 | 23.41 |
| Random | 748.56 | 702.81 | 1352.16 | 13.53 | 26.03 |
| **Diffusion Probability = 0.2** | | | | **Similarity Percentage = 0.7** | |
| **Method** | **Added** | **Random** | **Total** | **M1** | **M2** |
| Betweenness Centrality | 724.74 | 679.14 | 1348.26 | 11.27 | 22.38 |
| PageRank Centrality | 679.02 | 634.01 | 1327.97 | 11.29 | 23.66 |
| IM Rank | 762.18 | 723.51 | 1414.09 | 9.11 | 17.81 |
| Static Greedy | 803.04 | 762.5 | 1496.53 | 9.02 | 17.72 |
| Random | 745.66 | 694.53 | 1363.7 | 12.49 | 24.53 |

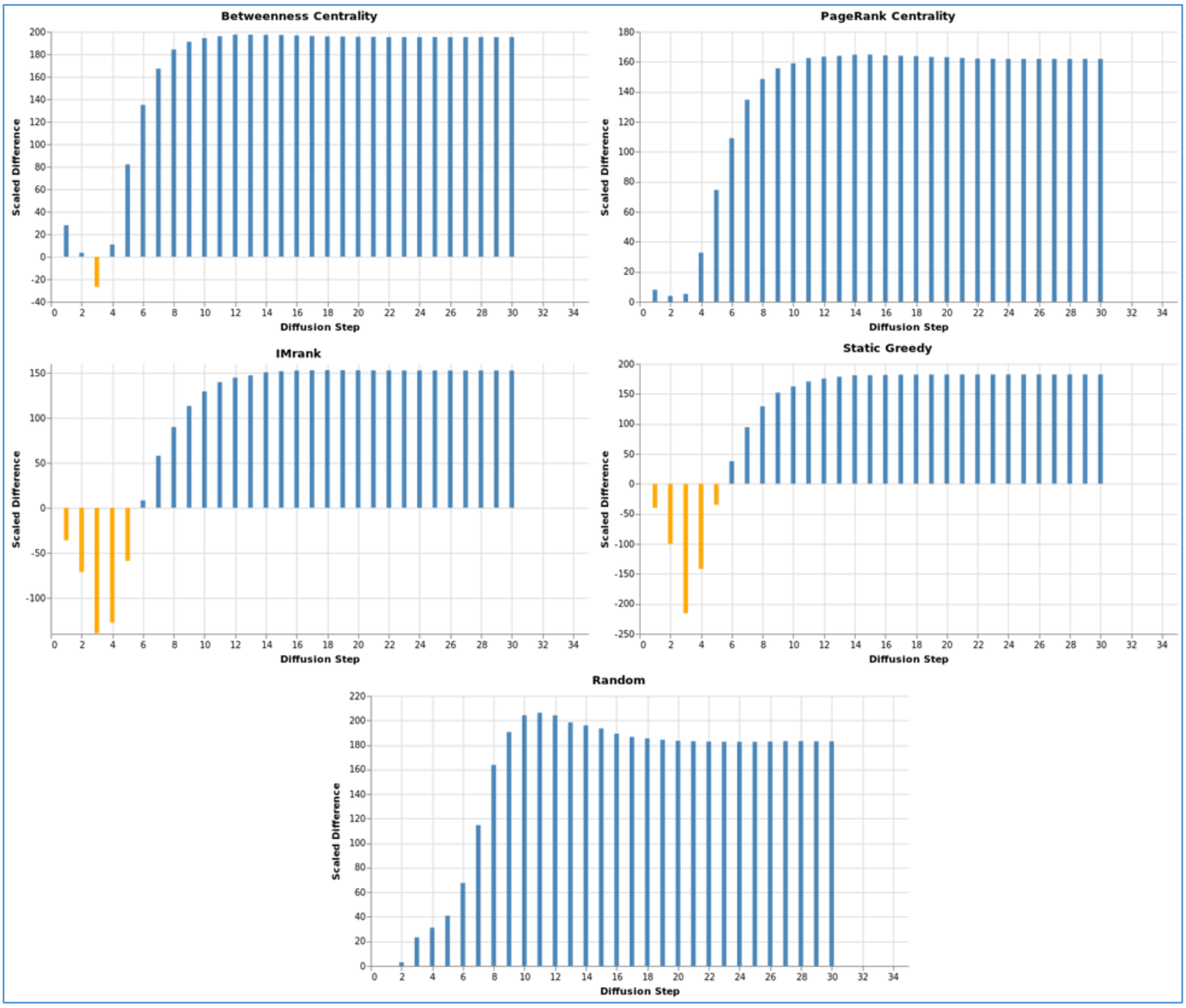

Figure 6. The trend at a similarity of 0.75 and a diffusion probability of 0.2

Table 10. The results of the first dataset with a diffusion probability of 0.15

| **Diffusion Probability = 0.15** | | | | **Similarity Percentage = 0.9** | |
|---|---|---|---|---|---|
| **Method** | **Added** | **Random** | **Total** | **M1** | **M2** |
| Betweenness Centrality | 472.97 | 463.14 | 975.36 | 10.07 | 21.22 |
| PageRank Centrality | 409.71 | 395.79 | 926.79 | 15.01 | 35.17 |
| IM Rank | 478.41 | 475.94 | 994.27 | 2.48 | 5.189 |
| Static Greedy | 509.72 | 504.32 | 1056.31 | 5.11 | 10.70 |
| Random | 461.64 | 440.8 | 908.13 | 22.94 | 47.27 |
| **Diffusion Probability = 0.15** | | | | **Similarity Percentage = 0.85** | |
| **Method** | **Added** | **Random** | **Total** | **M1** | **M2** |
| Betweenness Centrality | 449.28 | 430.48 | 963.71 | 13.00 | 29.11 |
| PageRank Centrality | 397.88 | 372.3 | 933.34 | 18.27 | 45.80 |
| IM Rank | 464.98 | 456.93 | 991.14 | 5.41 | 11.74 |
| Static Greedy | 496.7 | 478.4 | 1051.77 | 11.59 | 25.50 |
| Random | 428.77 | 400.46 | 886.93 | 21.27 | 47.12 |
| **Diffusion Probability = 0.15** | | | | **Similarity Percentage = 0.8** | |
| **Method** | **Added** | **Random** | **Total** | **M1** | **M2** |
| Betweenness Centrality | 453.41 | 416.66 | 968.97 | 18.96 | 44.10 |
| PageRank Centrality | 396.9 | 366.09 | 922.15 | 16.70 | 42.079 |
| IM Rank | 459.69 | 450.34 | 994.45 | 4.70 | 10.38 |
| Static Greedy | 487.85 | 472.53 | 1058.38 | 7.23 | 16.21 |
| Random | 424.1 | 398.57 | 879.31 | 14.51 | 32.02 |
| **Diffusion Probability = 0.15** | | | | **Similarity Percentage = 0.75** | |
| **Method** | **Added** | **Random** | **Total** | **M1** | **M2** |
| Betweenness Centrality | 427.44 | 413.62 | 966.4 | 5.72 | 13.36 |
| PageRank Centrality | 374.74 | 349.17 | 924.65 | 11.061 | 29.29 |
| IM Rank | 426.16 | 432.87 | 993.96 | -2.70 | -6.20 |
| Static Greedy | 452.59 | 450.19 | 1040.11 | 0.92 | 2.13 |
| Random | 382.93 | 372.15 | 872.03 | 4.94 | 11.58 |
| **Diffusion Probability = 0.15** | | | | **Similarity Percentage = 0.7** | |
| **Method** | **Added** | **Random** | **Total** | **M1** | **M2** |
| Betweenness Centrality | 406.78 | 392.03 | 963.66 | 5.10 | 12.54 |
| PageRank Centrality | 359.26 | 332.64 | 927.46 | 9.56 | 26.67 |
| IM Rank | 417.42 | 415.35 | 988.45 | 0.69 | 1.66 |
| Static Greedy | 440.89 | 430.79 | 1039.96 | 3.23 | 7.81 |
| Random | 404.84 | 375.76 | 888.84 | 10.90 | 25.79 |

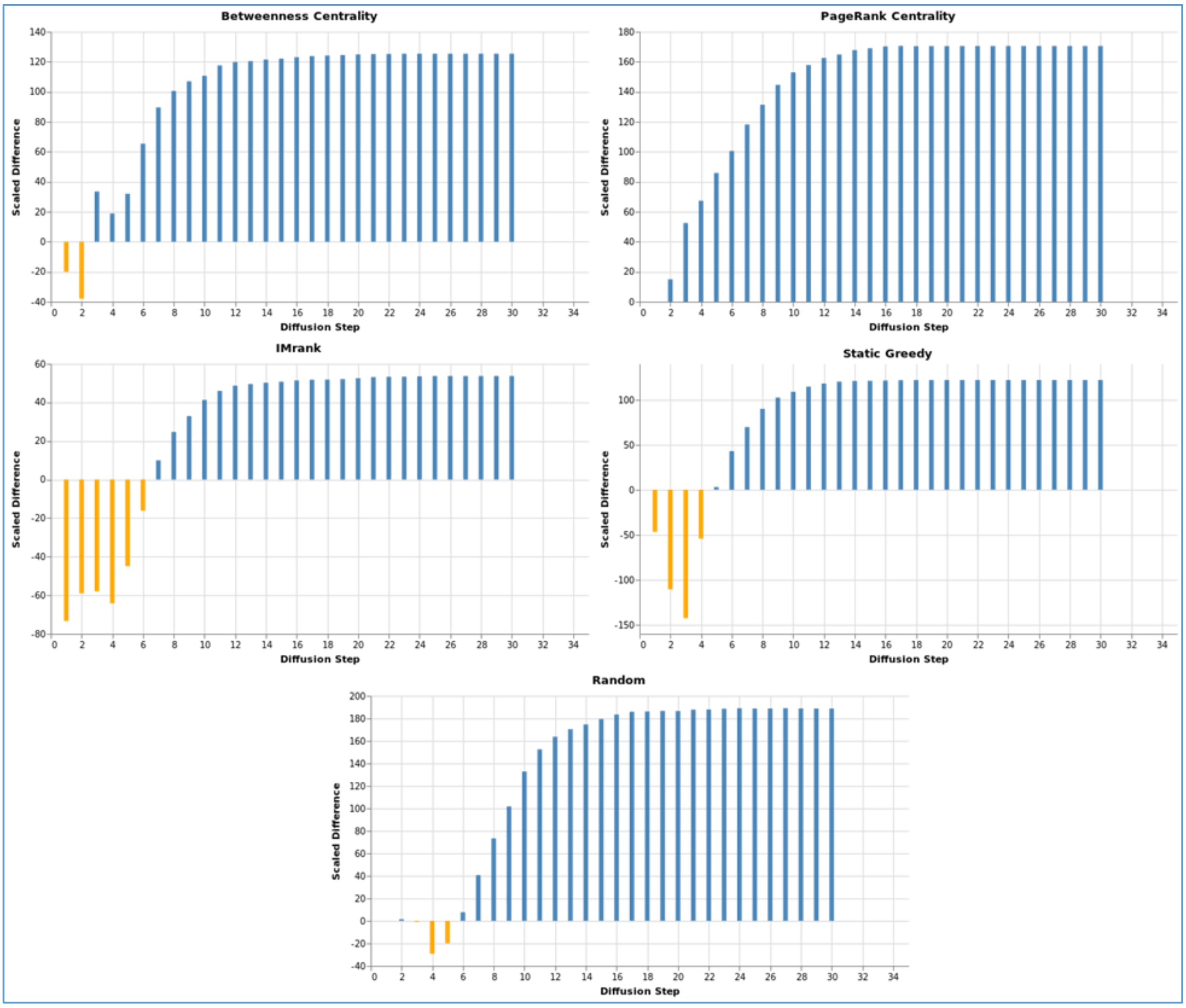

Figure 7. The trend at a similarity of 0.85 and a diffusion probability of 0.15

## 4.1 The Results of the Second Dataset

Table 11. The results of the second dataset with a diffusion probability of 0.25

| **Diffusion Probability = 0.25** | | | | **Similarity Percentage = 0.9** | |
|---|---|---|---|---|---|
| **Method** | **Added** | **Random** | **Total** | **M1** | **M2** |
| Betweenness Centrality | 2465.04 | 2447.58 | 3844.83 | 4.54 | 7.13 |
| PageRank Centrality | 2468.88 | 2452.71 | 3851.76 | 4.19 | 6.59 |
| IM Rank | 2502.23 | 2484.37 | 3878.92 | 4.60 | 7.18 |
| Static Greedy | 2552.75 | 2509.81 | 3943.65 | 10.88 | 17.10 |
| Random | 2587.48 | 2568.1 | 3970.44 | 4.88 | 7.54 |
| **Diffusion Probability = 0.25** | | | | **Similarity Percentage = 0.85** | |
| **Method** | **Added** | **Random** | **Total** | **M1** | **M2** |
| Betweenness Centrality | 2478.78 | 2424.67 | 3848 | 9.37 | 14.87 |
| PageRank Centrality | 2488.12 | 2430.01 | 3856.87 | 10.04 | 15.94 |
| IM Rank | 2506.77 | 2456.11 | 3872.28 | 8.72 | 13.75 |
| Static Greedy | 2546.41 | 2489.09 | 3936.53 | 9.70 | 15.35 |
| Random | 2581.17 | 2539.36 | 3948.08 | 7.059 | 10.97 |
| **Diffusion Probability = 0.25** | | | | **Similarity Percentage = 0.8** | |
| **Method** | **Added** | **Random** | **Total** | **M1** | **M2** |
| Betweenness Centrality | 2448.97 | 2405.4 | 3851.46 | 5.65 | 9.05 |
| PageRank Centrality | 2449.17 | 2398.91 | 3849.68 | 6.52 | 10.47 |
| IM Rank | 2472.53 | 2425.58 | 3855.26 | 6.08 | 9.67 |
| Static Greedy | 2523.86 | 2476.38 | 3944.1 | 6.01 | 9.58 |
| Random | 2563.74 | 2507.74 | 3951.78 | 7.08 | 11.165 |
| **Diffusion Probability = 0.25** | | | | **Similarity Percentage = 0.75** | |
| **Method** | **Added** | **Random** | **Total** | **M1** | **M2** |
| Betweenness Centrality | 2455.67 | 2380.11 | 3847.94 | 7.85 | 12.69 |
| PageRank Centrality | 2465.83 | 2389.52 | 3852.94 | 7.92 | 12.77 |
| IM Rank | 2483.42 | 2411.88 | 3867.88 | 7.39 | 11.86 |
| Static Greedy | 2524.68 | 2440.36 | 3935.12 | 8.57 | 13.82 |
| Random | 2566.72 | 2504.41 | 3966.42 | 6.28 | 9.95 |
| **Diffusion Probability = 0.25** | | | | **Similarity Percentage = 0.7** | |
| **Method** | **Added** | **Random** | **Total** | **M1** | **M2** |
| Betweenness Centrality | 2455.65 | 2369.83 | 3851.91 | 7.42 | 12.07 |
| PageRank Centrality | 2456.14 | 2369.95 | 3860.7 | 7.44 | 12.12 |
| IM Rank | 2482.73 | 2385.38 | 3869.55 | 8.38 | 13.60 |
| Static Greedy | 2524.77 | 2414.51 | 3930.89 | 9.34 | 15.22 |
| Random | 2570.77 | 2485.15 | 3973.55 | 7.18 | 11.48 |

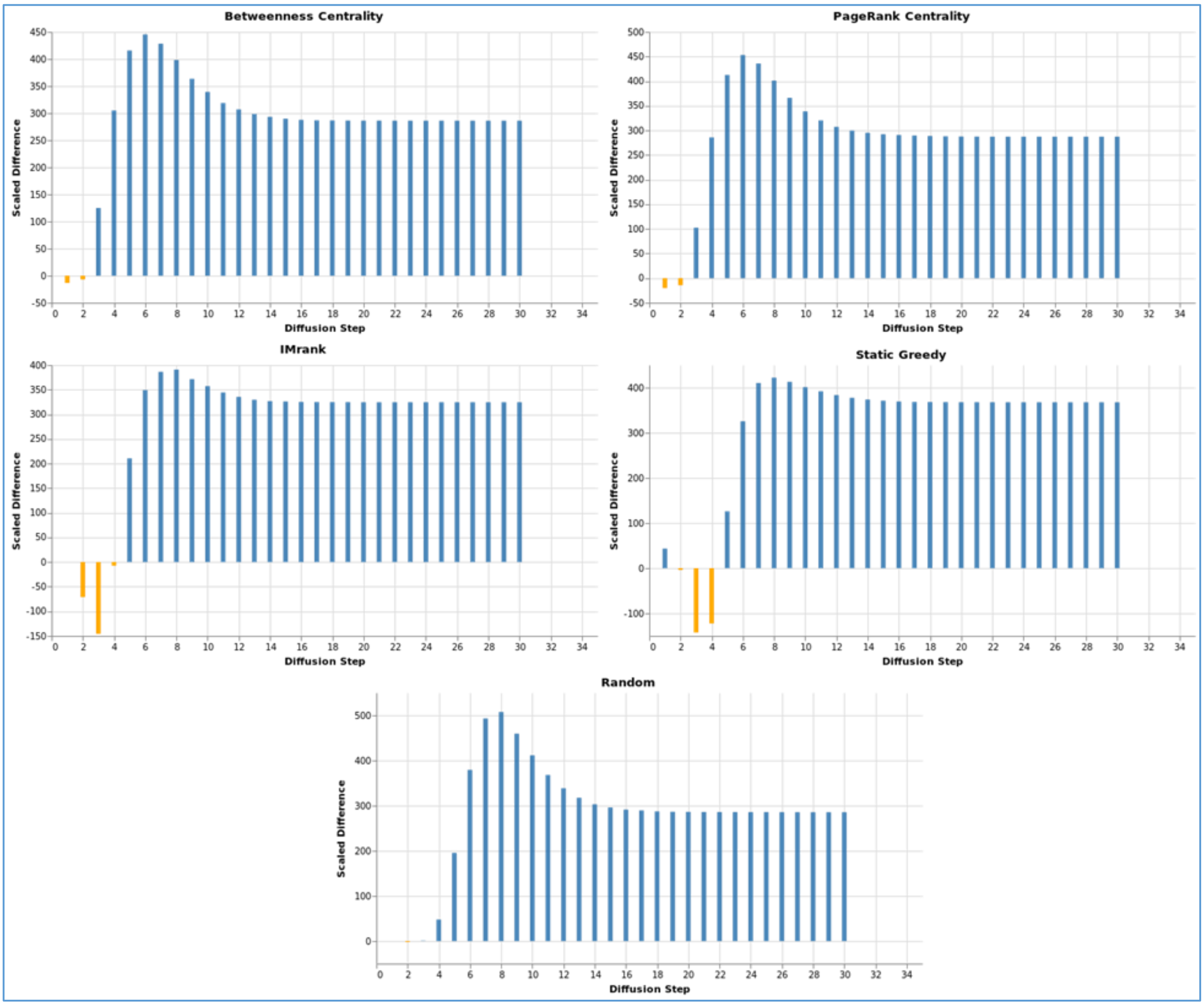

Figure 8. The trend at a similarity of 0.7 and a diffusion probability of 0.25

Table 12. The results of the second dataset with a diffusion probability of 0.2

| **Diffusion Probability = 0.2** | | | | **Similarity Percentage = 0.9** | |
|---|---|---|---|---|---|
| **Method** | **Total** | **Random** | **Total** | **M1** | **M2** |
| Betweenness Centrality | 1691.14 | 1648.54 | 2990.37 | 14.24 | 25.84 |
| PageRank Centrality | 1693.91 | 1662.04 | 3014.66 | 10.57 | 19.17 |
| IM Rank | 1738.7 | 1701.11 | 3018.99 | 12.45 | 22.09 |
| Static Greedy | 1793.47 | 1734.84 | 3123.04 | 18.77 | 33.79 |
| Random | 1782.93 | 1752.01 | 3099.06 | 9.97 | 17.64 |
| **Diffusion Probability = 0.2** | | | | **Similarity Percentage = 0.85** | |
| **Method** | **Total** | **Random** | **Total** | **M1** | **M2** |
| Betweenness Centrality | 1675.54 | 1626.08 | 3002.99 | 10.98 | 20.27 |
| PageRank Centrality | 1678.63 | 1626.18 | 3000.58 | 11.65 | 21.50 |
| IM Rank | 1717.35 | 1671.33 | 3034.43 | 10.11 | 18.35 |
| Static Greedy | 1779.62 | 1712.78 | 3134.12 | 14.21 | 26.01 |
| Random | 1783.7 | 1731.91 | 3104.82 | 11.12 | 19.93 |
| **Diffusion Probability = 0.2** | | | | **Similarity Percentage = 0.8** | |
| **Method** | **Total** | **Random** | **Total** | **M1** | **M2** |
| Betweenness Centrality | 1654.99 | 1590.59 | 2984.31 | 10.78 | 20.24 |
| PageRank Centrality | 1665.94 | 1597.09 | 3006.75 | 11.44 | 21.55 |
| IM Rank | 1703.01 | 1642.72 | 3036.81 | 9.92 | 18.35 |
| Static Greedy | 1769.09 | 1673.67 | 3133.4 | 15.22 | 28.50 |
| Random | 1754.78 | 1679.88 | 3093.07 | 12.10 | 22.29 |
| **Diffusion Probability = 0.2** | | | | **Similarity Percentage = 0.75** | |
| **Method** | **Total** | **Random** | **Total** | **M1** | **M2** |
| Betweenness Centrality | 1621.9 | 1550.04 | 2980.88 | 9.64 | 18.54 |
| PageRank Centrality | 1625.08 | 1559.8 | 3003.14 | 8.694 | 16.74 |
| IM Rank | 1674.93 | 1595.02 | 3020.95 | 10.58 | 20.03 |
| Static Greedy | 1728.78 | 1641.47 | 3138.74 | 11.12 | 21.27 |
| Random | 1720.21 | 1644.68 | 3085.25 | 9.79 | 18.36 |
| **Diffusion Probability = 0.2** | | | | **Similarity Percentage = 0.7** | |
| **Method** | **Total** | **Random** | **Total** | **M1** | **M2** |
| Betweenness Centrality | 1622.22 | 1521.68 | 2987.91 | 11.21 | 22.02 |
| PageRank Centrality | 1628.9 | 1528.48 | 2995.65 | 11.17 | 21.89 |
| IM Rank | 1668.46 | 1569.78 | 3027.09 | 10.86 | 20.95 |
| Static Greedy | 1708.24 | 1605.82 | 3132.96 | 10.89 | 21.26 |
| Random | 1729.65 | 1639.28 | 3108.2 | 9.69 | 18.37 |

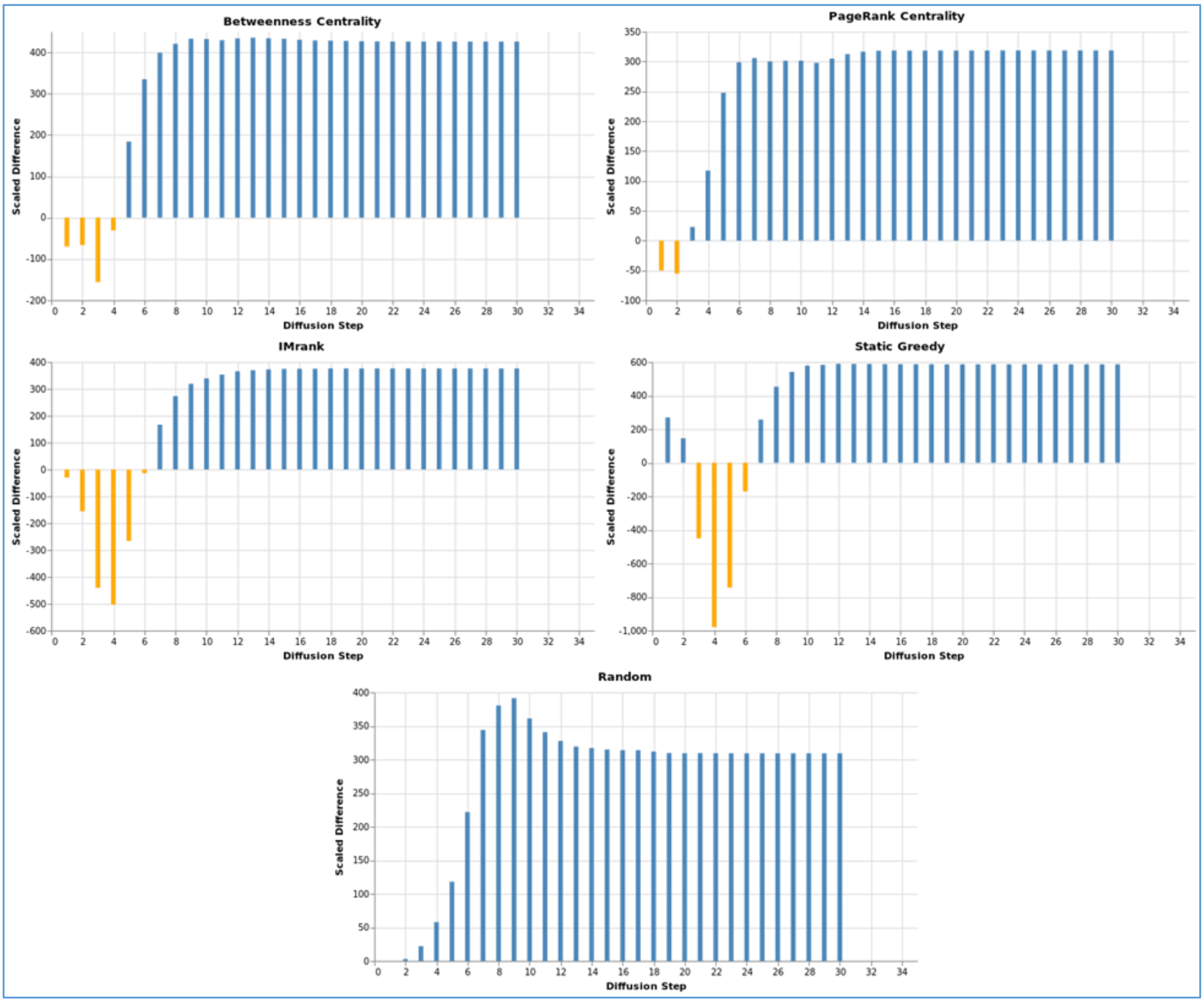

Figure 9. The trend at a similarity of 0.9 and a diffusion probability of 0.2

Table 13. The results of the second dataset with a diffusion probability of 0.15

| **Diffusion Probability = 0.15** | | | | **Similarity Percentage = 0.9** | |
|---|---|---|---|---|---|
| **Method** | **Total** | **Random** | **Total** | **M1** | **M2** |
| Betweenness Centrality | 932.66 | 908.75 | 2022.56 | 11.82 | 26.31 |
| PageRank Centrality | 926.83 | 903.81 | 2015.07 | 11.42 | 25.46 |
| IM Rank | 985.84 | 970.77 | 2057.78 | 7.32 | 15.52 |
| Static Greedy | 997.13 | 967.34 | 2071.22 | 14.38 | 30.79 |
| Random | 990.92 | 961.87 | 2081.99 | 13.95 | 30.20 |
| **Diffusion Probability = 0.15** | | | | **Similarity Percentage = 0.85** | |
| **Method** | **Total** | **Random** | **Total** | **M1** | **M2** |
| Betweenness Centrality | 902.52 | 870.65 | 2013.63 | 10.55 | 24.40 |
| PageRank Centrality | 883.31 | 863.57 | 2008.47 | 6.55 | 15.23 |
| IM Rank | 955.04 | 926.71 | 2063.69 | 9.15 | 20.38 |
| Static Greedy | 957.43 | 925.56 | 2059.64 | 10.31 | 22.95 |
| Random | 967.25 | 924.32 | 2083.39 | 13.73 | 30.96 |
| **Diffusion Probability = 0.15** | | | | **Similarity Percentage = 0.8** | |
| **Method** | **Total** | **Random** | **Total** | **M1** | **M2** |
| Betweenness Centrality | 865.93 | 840.61 | 2017.69 | 6.27 | 15.06 |
| PageRank Centrality | 858.57 | 839.86 | 2018.88 | 4.63 | 11.13 |
| IM Rank | 918.79 | 899.19 | 2071.84 | 4.73 | 10.89 |
| Static Greedy | 917.43 | 900.1 | 2062.7 | 4.20 | 9.62 |
| Random | 912.42 | 894.84 | 2077.39 | 4.23 | 9.82 |
| **Diffusion Probability = 0.15** | | | | **Similarity Percentage = 0.9** | |
| **Method** | **Total** | **Random** | **Total** | **M1** | **M2** |
| Betweenness Centrality | 836.99 | 795.29 | 2028.05 | 8.22 | 20.97 |
| PageRank Centrality | 841.47 | 802.8 | 2026.48 | 7.63 | 19.26 |
| IM Rank | 895.45 | 863.13 | 2069.67 | 6.24 | 14.97 |
| Static Greedy | 890.85 | 864.17 | 2076.03 | 5.14 | 12.34 |
| Random | 887.64 | 844.26 | 2084.53 | 8.32 | 20.55 |
| **Diffusion probability = 0.15** | | | | **Percentage of similarity = 0.7** | |
| **Method** | **Total** | **Random** | **Total** | **M1** | **M2** |
| Betweenness Centrality | 848.23 | 763.85 | 2019 | 13.93 | 36.82 |
| PageRank Centrality | 850.63 | 763.95 | 2011.18 | 14.36 | 37.82 |
| IM Rank | 895.74 | 823.34 | 2063.47 | 11.69 | 29.31 |
| Static Greedy | 906.17 | 828.79 | 2082.16 | 12.38 | 31.12 |
| Random | 910.38 | 796.25 | 2071.42 | 18.36 | 47.77 |

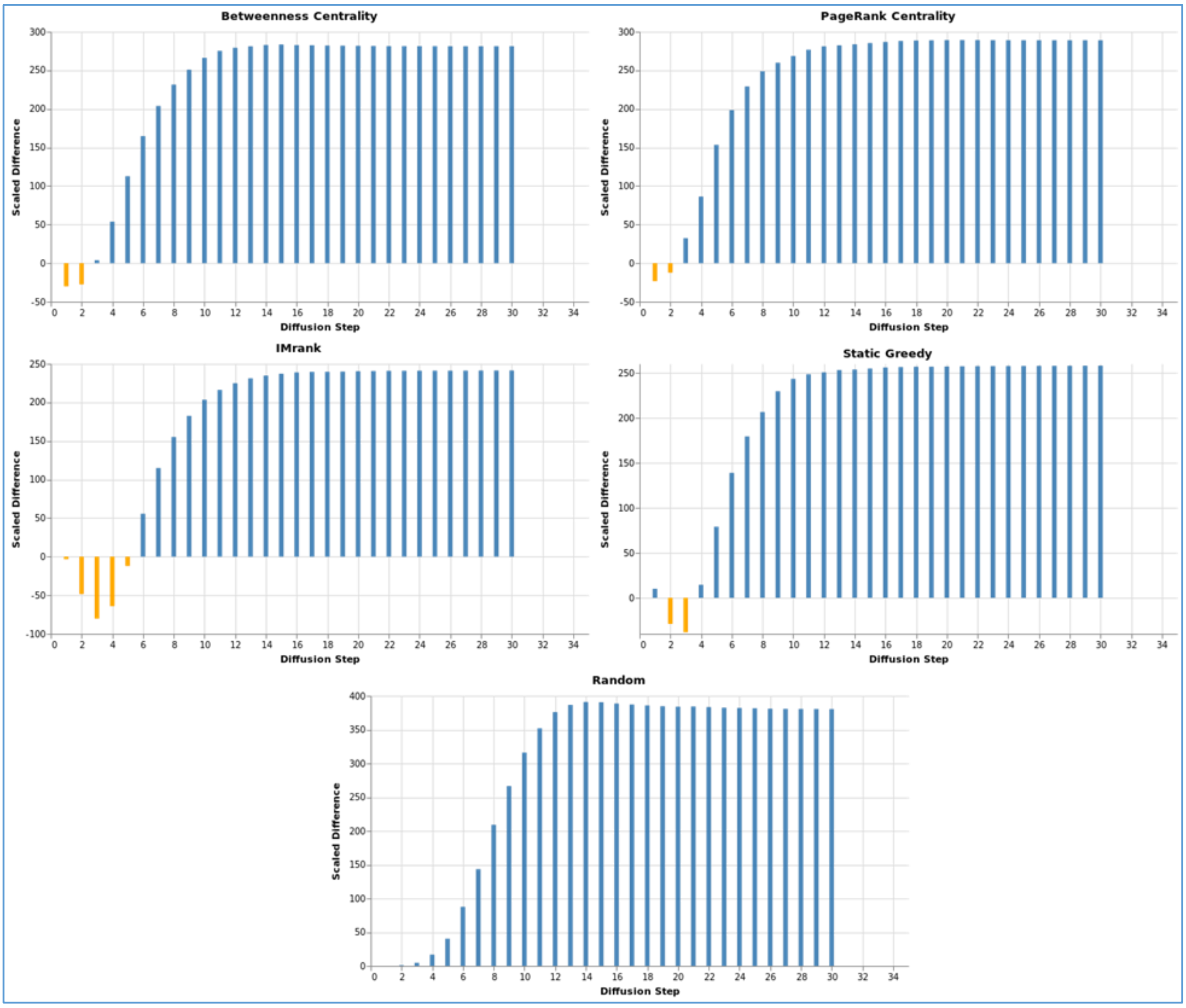

Figure 10. The trend at a similarity of 0.7 and a diffusion probability of 0.15

## 4.2 Results Analysis

The results were again sorted based on the best results, and Tables 14 and 15 were obtained.

Table 14. The best results of the first dataset

| Method | Diffusion Probability | Similarity Percentage | M1 | M2 |
|---|---|---|---|---|
| Betweenness Centrality | 0.15 | 0.8 | 18.96 | 44.10 |
| PageRank Centrality | 0.15 | 0.85 | 18.27 | 45.80 |
| IM Rank | 0.25 | 0.75 | 10.75 | 20.57 |
| Static Greedy | 0.25 | 0.75 | 12.07 | 23.41 |
| Random | 0.15 | 0.85 | 21.27 | 47.12 |

Table 15. The best results of the second dataset

| Method | Diffusion Probability | Similarity Percentage | M1 | M2 |
|---|---|---|---|---|
| Betweenness Centrality | 0.2 | 0.9 | 14.24 | 25.84 |
| PageRank Centrality | 0.15 | 0.7 | 14.36 | 37.82 |
| IM Rank | 0.2 | 0.9 | 12.45 | 22.09 |
| Static Greedy | 0.2 | 0.9 | 18.77 | 33.79 |
| Random | 0.15 | 0.7 | 18.36 | 47.77 |

The green cells in Tables 8 to 13 show the best results of each of the five investigated algorithms for different values removed from the original graph and the best results of each of the methods on different diffusion coefficients. Dark green cells are the priority, and light green cells are the next. According to the tables, the majority of the best results were obtained at lower diffusion coefficients because most of the nodes were activated at high diffusion coefficients regardless of the algorithm. The algorithm can demonstrate its superiority in low diffusion coefficients. There was no significant difference between the results of different values removed from the graph. This case can be an advantage of the proposed method because the results had a kind of consistency between the different values removed from the graph. According to the trends drawn in figures 5 to 10, the IM performed weaker than other methods in the initial steps but achieved the best results in the long term. This could be attributed to the foresight in the proposed IM algorithm. This means that these algorithms look for nodes that have maximum influence in the long term and during the diffusion of a message, not just in a subgraph of the original graph.

## 5- Conclusion and Future Studies

Social networks (SNs) are very popular nowadays because they are very cheap communication channels for establishing social communication between people. So, many websites have included SN facilities and infrastructures. SNs are social structures consisting of people and relationships between them and play an important role in spreading and transferring information between people today. The members of an SN share different topics and diffuse different information among themselves. In these networks, one person's opinions, thoughts, and beliefs can change those of other members. So, SN analysts have paid a lot of attention to expanding influence and identifying influential people. Some people are more influential and some less. One of the most important problems in influence optimization is to find a small subset of social people so that, by activating them, the information of the largest number of people from the network is influenced under a diffusion model. The selection of influential nodes is crucial for enhancing knowledge and adopting behavior in a network because they can influence other nodes. A better understanding of the structure and behavior of the network can be achieved by using the highlighted nodes. The nodes that need more attention and investment to perform a specific task can be found by identifying the influential people in the network from the perspective of various parameters. This study was conducted to develop a new IM method in real social networks with graphs with limited visibility. The innovation of the study is the operationalization of the previous IM methods in real environments with limited visibility. To this end, a link prediction step was performed before the execution of the IM algorithm. This method made it possible to recover some of the lost dynamics in the network. Moreover, an experimental framework was provided to examine the performance of the method during the tests performed on real datasets. The results suggested the improvement of the possible efficiency of the proposed method in real-world problems. In summary, the following were performed in this study:

- Investigating link prediction capabilities of a graph generation model exponential random graph model (ERGM);
- Reviewing related studies and mentioning their advantages and noteworthy issues;
- Matching real-world graphs onto an ERGM;
- Performing a link prediction using the matched ERGM;
- Performing a step of IM on the ERGM output graph;
- Implementing all the above;
- Providing a framework to test the capabilities of the proposed method;
- Conducting extensive tests on real-world datasets that showed the capabilities of the proposed method.

As could be seen, the learning stage of the link prediction model was completely independent of the IM in the proposed method. Future studies are recommended to investigate the feasibility of simultaneous training of these two parts. The training part of the link prediction model can be changed in such a way that it receives direct feedback from the IM results and the link prediction method training is provided exclusively for IM. Future studies are also recommended to consider the contents of network nodes. In addition to the graphs showing the links, the real-world networks have many other data which, if combined with the current models, can add more power to the proposed model. The use of other methods of link prediction and graph generation, such as those introduced in [76], are important research branches for future work.